\documentclass{aa} 

\usepackage{array}
\usepackage{graphicx}
\usepackage{amsmath}
\usepackage{longtable}
\usepackage{setspace}

\usepackage{txfonts}
\usepackage{soul}
\usepackage{booktabs}
\usepackage{pdfpages}
\usepackage{url}

\usepackage[colorlinks=true,linkcolor=blue, citecolor=blue]{hyperref}
\newcommand\Sun{\sun}%
\providecommand{\tjdtdb}{\ensuremath{\rm {TJD_{TDB}}}}

\providecommand{\rsun}{\ensuremath{\,R_\Sun}}
\providecommand{\lsun}{\ensuremath{\,L_\Sun}}

\providecommand{\me}{\ensuremath{\,M_{\rm E}}}
\providecommand{\re}{\ensuremath{\,R_{\rm E}}}

\usepackage{upgreek}
\usepackage{float}

\begin{document}

\title{Hot Rocks Survey VI: An anomalously hot dayside for the rocky planet GJ~357~b}
\author{
    Merlin Zgraggen\inst{1,2}\corrauth{mzgragge@ethz.ch}
    \and Brice-Olivier Demory\inst{1,3,4}\email{brice.demory@unibe.ch}
    \and Hannah Diamond-Lowe\inst{5}\email{hdiamondlowe@stsci.edu}
    \and João M. Mendonça\inst{6,7}\email{j.mendonca@soton.ac.uk}
    \and Erik Meier Valdés\inst{8}\email{erik.meiervaldes@physics.ox.ac.uk}
    \and Mark Fortune\inst{9}\email{mark.fortune@mq.edu.au}
    \and Kathryn D. Jones\inst{1}\email{kathryndjones@hotmail.co.uk}
    \and Daniel Kitzmann\inst{3}\email{daniel.kitzmann@unibe.ch}
    \and Millie Smith\inst{6}\email{E.R.Smith@soton.ac.uk}
    \and Natalie H. Allen\inst{10}\email{natalie.allen@umontreal.ca}
    \and Prune C. August\inst{11}\email{prua@space.dtu.dk}
    \and Amélie Gressier\inst{12}\email{amelie.gressier@umontreal.ca}
    \and Måns Holmberg\inst{5}\email{mholmberg@stsci.edu}
    \and Jegug Ih\inst{5}\email{jih@stsci.edu}
    \and Lars A. Buchhave\inst{11}\email{buchhave@space.dtu.dk}
    \and Néstor Espinoza\inst{5,10}\email{nespinoza@stsci.edu}
    \and Neale P. Gibson\inst{13}\email{n.gibson@tcd.ie}
    \and Kevin Heng\inst{14,15,16}\email{Kevin.Heng@physik.lmu.de}
}

\institute{
    Centre for Space and Habitability,
    University of Bern,
    Gesellschaftsstrasse 6,
    3012 Bern,
    Switzerland
    \and
    Institute for Particle Physics and Astrophysics,
    Department of Physics,
    ETH Zurich,
    Wolfgang-Pauli-Strasse 27,
    8093 Zurich,
    Switzerland
    \and
    Space Research and Planetary Sciences,
    Physics Institute,
    University of Bern,
    Gesellschaftsstrasse 6,
    3012 Bern,
    Switzerland
    \and
    ARTORG Center for Biomedical Engineering Research,
    University of Bern,
    Switzerland
    \and
    Space Telescope Science Institute,
    3700 San Martin Drive,
    Baltimore, MD 21218,
    USA
    \and
    School of Physics and Astronomy,
    University of Southampton,
    Highfield,
    Southampton SO17 1BJ,
    UK
    \and
    School of Ocean and Earth Science,
    University of Southampton,
    Southampton SO14 3ZH,
    UK
    \and
    Department of Physics,
    University of Oxford,
    Keble Road,
    Oxford OX1 3RH,
    UK
    \and
    School of Mathematical \& Physical Sciences,
    Macquarie University,
    12 Wally's Walk,
    Macquarie Park, NSW 2113,
    Australia
    \and
    Department of Physics and Astronomy,
    Johns Hopkins University,
    3400 N. Charles Street,
    Baltimore, MD 21218,
    USA
    \and
    DTU Space,
    Technical University of Denmark,
    Elektrovej 328,
    2800 Kgs. Lyngby,
    Denmark
    \and
    Department of Physics and Trottier Institute for Research on Exoplanets,
    Universit\'{e} de Montr\'{e}al,
    Montr\'{e}al, QC,
    Canada
    \and
    School of Physics,
    Trinity College Dublin,
    University of Dublin,
    Dublin 2,
    Ireland
    \and
    Faculty of Physics,
    Ludwig Maximilian University,
    Scheinerstr. 1,
    D-81679 Munich,
    Germany
    \and
    Munich Center for Geoastronomy,
    Ludwig Maximilian University,
    Theresienstrasse 41,
    D-80333 Munich,
    Germany
    \and
    Department of Physics \& Astronomy,
    University College London,
    Gower Street,
    London WC1E 6BT,
    United Kingdom
}


\authorrunning{M. Zgraggen et al.}
 
\date{}
\abstract
    {As part of the \textit{JWST Hot Rocks Survey}, we analyzed a JWST/MIRI F1500W secondary-eclipse observation of GJ~357~b, a rocky exoplanet orbiting a nearby M dwarf. We also performed a new global analysis of the GJ~357 system, combining \textit{TESS} photometry, a JWST/NIRSpec transit observation, the MIRI secondary eclipse, radial velocities, and \textit{Gaia} parallax measurements to obtain a self-consistent set of stellar and planetary parameters. We measure an occultation depth of $200.5 \pm 12.7\,\mathrm{ppm}$, corresponding to a dayside brightness temperature at $15\,\mu\mathrm{m}$ of $T_{\mathrm{b},15\,\mu\mathrm{m}}=923^{+39}_{-38}\,\mathrm{K}$. The inferred brightness temperature substantially exceeds the zero-Bond-albedo, no-redistribution, disk-integrated bolometric dayside reference temperature of $686 \pm 14\,\mathrm{K}$. The measured eclipse depth differs from the tested zero-albedo, grey airless-surface model by $5.2\sigma$ and is also inconsistent with the high-mean-molecular-weight atmospheric models considered here, including pure CO$_2$, pure H$_2$O, and N$_2$ with 100~ppm CO$_2$. Possible explanations for the excess emission in the MIRI/F1500W band include wavelength-dependent surface emissivity, localized hot regions, or atmospheric temperature structures and opacity sources not included in the tested models. Follow-up mid-infrared eclipse spectroscopy would constrain the shape of the dayside emission spectrum and help test the proposed surface and atmospheric explanations.}


\maketitle

\nolinenumbers
%

\section{Introduction}

M dwarfs are favourable targets for detecting and characterising small exoplanets due to their small radii (0.1–0.6\,R$_\odot$), which increase transit and secondary eclipse depths, and their low masses, which enhance radial velocity signals \citep{Charbonneau2007,Tarter2007}. At the same time, their long pre-main-sequence phase and high XUV output may drive significant atmospheric escape \citep[e.g.][]{Shields2016,Lammer2014,Tian2009,Luger2015,Owen2016,Owen_2019,Krissansen-Totton2023}, and stellar winds could contribute to atmospheric erosion over Gyr timescales \citep{Garcia-Sage2017,Dong2018,Diamond-Lowe2021}.

The \textbf{JWST Hot Rocks Survey}, conducted under JWST GO~3730 \citep{Diamond-Lowe2023}, uses JWST/MIRI secondary eclipses (sometimes called occultations) using photometric observations at $15\,\mu\mathrm{m}$ to measure the dayside thermal emission of short-period rocky planets. By comparing the observed brightness temperatures to bare-rock and high-mean-molecular-weight (high-$\mu$) atmospheric models, the survey tests whether these planets retain substantial atmospheres or are potentially airless. Early survey results already show a diverse picture. LHS~1478\,b \citep{August2025} shows a possible shallow eclipse in the first visit, while the second visit was affected by stronger systematics and did not yield a significant detection, leaving the atmospheric interpretation uncertain. TOI-1468\,b \citep{Meier2025} exhibits a moderate excess in dayside thermal emission relative to the bare-rock prediction, although the result remains consistent with the bare-rock case within the uncertainties. LHS~1140\,c \citep{Fortune2025} instead shows a robust deep eclipse consistent with a low-albedo bare-rock surface and inefficient heat redistribution. LTT~3780\,b \citep{Allen2025} similarly appears consistent with an atmosphere-free low-albedo dayside, while still allowing some thin or spectrally inactive atmospheres.
GJ~3473\,b \citep{Holmberg2026} illustrates the remaining degeneracy of single-band F1500W photometry: both bare-rock and atmospheric interpretations remain possible, although thick CO$_2$ atmospheres are disfavoured. 

GJ~357~b is a transiting rocky exoplanet orbiting the nearby M2.5\,V star GJ~357 \citep{Luque2019}. From the global analysis presented in Section~\ref{subsec:globalfit}, we derive a planetary mass of $2.05 \pm 0.43$\,M$_\oplus$, a planetary radius of $1.152^{+0.034}_{-0.032}$\,R$_\oplus$, and a stellar mass of $0.3455^{+0.0078}_{-0.0077}$\,M$_\odot$. The system also hosts two non-transiting planets, GJ~357\,c and d, detected through radial velocities \citep{Luque2019}. In addition to \emph{TESS} photometry and radial velocities, GJ~357\,b has been observed in transmission with JWST/NIRISS SOSS \citep{Taylor_2025} and JWST/NIRSpec G395H \citep{AdamsRedai2025}. Both transmission spectra are featureless at the tens-of-ppm level, ruling out extended H/He atmospheres and clear, metal-poor secondary atmospheres. High-$\mu$ atmospheres with small scale heights and surface-dominated emission remain consistent with the data.

In this paper we present the JWST/MIRI F1500W secondary eclipse observation of GJ~357\,b and interpret the measured dayside emission in the context of bare-surface and atmospheric forward models. Section~\ref{sec:obs} describes the observations. Section~\ref{sec:datareduction} presents the data reduction, global fit, eclipse analysis, and robustness tests. Section~\ref{sec:results} presents the brightness-temperature inference and comparison with forward models. Section~\ref{sec:discussion} discusses possible physical explanations for the high measured flux and outlines the observations needed to distinguish between them. Section~\ref{sec:conclusion} summarizes our conclusions.

\section{Observations}\label{sec:obs}

\subsection{JWST/MIRI secondary eclipse observations}

We observed one secondary eclipse of GJ 357\,b using the Mid-Infrared Instrument \citep[MIRI;][]{Rieke2015} aboard the \textit{James Webb Space Telescope} \citep{Gardner2006,Rigby2023}, as part of the \textit{Hot Rocks Survey} (GO \#3730, PI: Diamond-Lowe, Co-PI: Mendon\c{c}a). The dataset was obtained in imaging mode with the F1500W filter on 11 May 2024, capturing the planet’s thermal emission around $15\,\mu\mathrm{m}$. This wavelength range is well suited for constraining the dayside brightness temperature of terrestrial exoplanets and potentially identifying deviations from a simple bare-rock thermal spectrum caused by atmospheric opacity.

The observations were executed in FASTR1 readout mode using the SUB64 subarray to avoid saturation \citep[see JWST/MIRI Instrument Handbook;][]{MIRIhandbook}. Each integration consisted of 22 groups, with a total of 8125 integrations, yielding a total observing duration of approximately 3.9 hours. A timing constraint on the phase ensured coverage of the predicted secondary eclipse with baseline before and after the eclipse. These constraints were calculated from publicly available TESS and RV measurements from \citet{Luque2019} and our own analysis. Baseline coverage before and after the eclipse was roughly equal, extending at least one eclipse duration on both sides to robustly establish the out-of-eclipse flux.

To account for known MIRI-specific systematics such as the time-dependent detector settling effect (e.g., \citealt{Powell2024, Zhang2024, Bell2024}), we included a 30-minute settling time before ingress. The ramp is characterized by a non-linear drift in detector response---typically a decay or rise in flux over the first few hundred seconds---and is believed to result from trap-filling and memory effects in the MIRI detector \citep{Dicken2024}. The effect is shown to be highly correlated with stellar flux \citep{Connors2025}.

Unlike several other targets in the Hot Rocks Survey, GJ~357\,b has no known close-in transiting sibling planets whose transits or occultations could overlap the observing window, though there are two known non-transiting planets on wider orbits detected in RVs \citep{Luque2019}. For planning purposes, the eclipse was scheduled assuming a nearly circular orbit. The host star's long rotation period of $\sim$78 days \citep{Oddo2023} further reduces the likelihood of rotationally modulated stellar variability affecting the short eclipse visit.

\section{Data reduction and analysis}
\label{sec:datareduction}

\subsection{Fiducial MIRI reduction} \label{reduction_occultation_fiducial}

\noindent\textbf{\textit{Note:}} \textit{The observation took place during an intense period of solar activity, including a series of strong solar flares and geomagnetic storms from 10--13 May 2024. Notably, the peak geomagnetic storm onset occurred on 11 May 2024, the same day as this observation, potentially influencing space-weather conditions \citep{Weiler2024,rodriguez2025,Liu_2024}
.}
\vspace{0.5em}

The \texttt{rateints} were reduced using the JWST pipeline version 1.16.0 steps specialized for MIRI TSO \citep{bushouse_2022_7229890} in the following order starting from the \texttt{uncal} datasets: \texttt{group\_scale}, \texttt{dq\_init}, \texttt{emicorr}, \texttt{saturation}, \texttt{reset}, \texttt{linearity}, \texttt{dark\_current}, \texttt{refpix}, \texttt{jump}, \texttt{ramp\_fit}, and \texttt{gain\_scale}. Due to the increased solar activity period, including a series of strong solar flares during the observation, we did not skip the \texttt{jump} step. The jump-detection threshold was set to 4$\sigma$, selected by minimizing the median absolute deviation (MAD) of a linear fit to the out-of-occultation light curve after the \texttt{ramp\_fit} step.

As noted by \citet{Morrison_2023}, the last group, group 22, showed an overall offset from the approximately linear ramp defined by groups 1--21. This was visible in individual pixels and in the summed pixel array, including pixels not masked by the saturation step. The last group was therefore systematically disregarded during the \texttt{ramp\_fit} step. Following this, the \texttt{assign\_wcs}, \texttt{flat\_field}, \texttt{photom}, and \texttt{outlier\_detection} steps were performed with standard settings.

\begin{figure}
    \centering
    \includegraphics[width=0.5\textwidth]{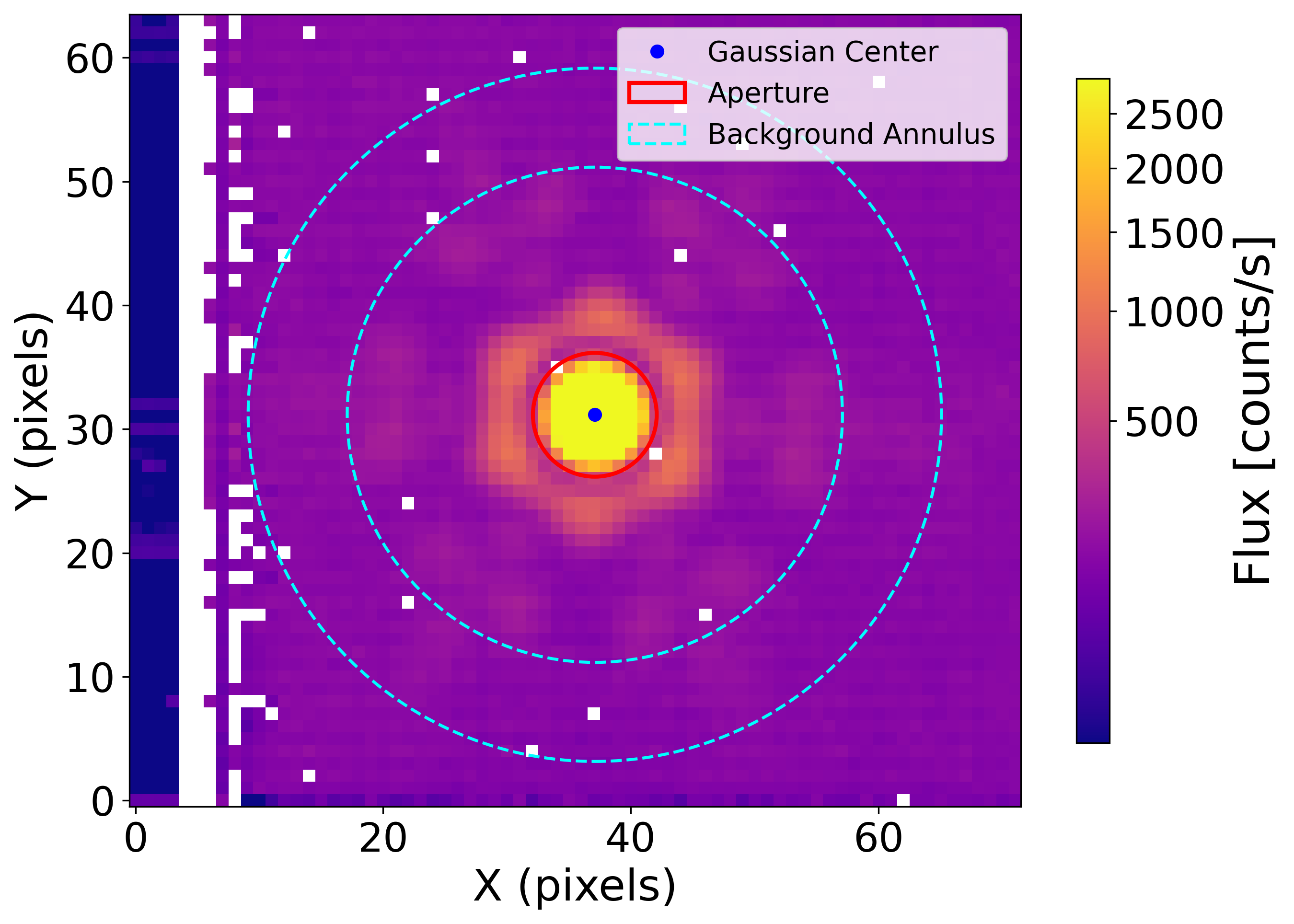}
    \caption{Example MIRI/F1500W SUB64 image from integration 100. The adopted 5-pixel radius aperture is shown together with the background annulus used for subtraction. The extended F1500W PSF structure motivates the aperture-optimization test shown in Fig.~\ref{fig:GJ357b_occultation_plots}.}
    \label{fig:Background_old_v3_miri_integration_100_aperture_plot}
\end{figure}

To extract the light curve, we used a custom Python pipeline to determine the PSF centroid in each integration via a two-dimensional Gaussian fit. Aperture photometry was performed using the \texttt{photutils} package within the Astropy ecosystem \citep{astropy2018}, employing a circular, pixel-weighted aperture in which edge pixels contribute according to their fractional overlap with the aperture. Background levels were estimated using a circular annulus with radii of 20--28 pixels and subtracted from the aperture flux. Pixels flagged with bad data-quality values were replaced using the fitted Gaussian PSF model. Uncertainties were propagated from photon noise and background subtraction using the median background level measured in the annulus for each integration.

\begin{figure}
\centering
\includegraphics[width=0.5\textwidth]{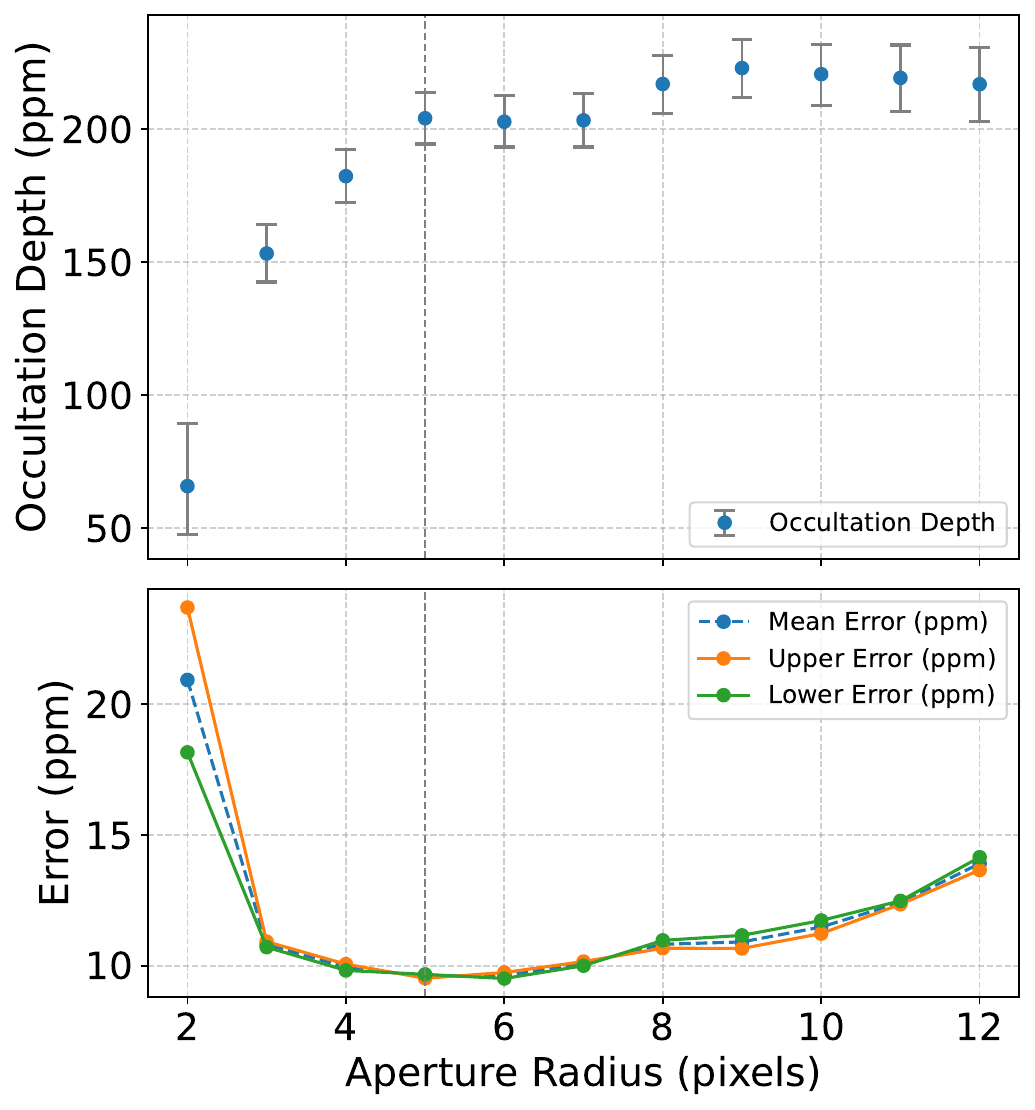}
\caption{Occultation depth (top) and corresponding uncertainty (bottom) as a function of aperture radius for GJ~357\,b. For these linear fits, the first 30 minutes of the observation were excluded to remove the initial time-dependent ramp. The inferred depth stabilizes at approximately $r=5$ pixels, where the uncertainty also reaches a minimum; we therefore adopt a 5-pixel aperture radius for the fiducial analysis.}
\label{fig:GJ357b_occultation_plots}
\end{figure}

The MIRI/Imaging SUB64 subarray is $64\times72$ pixels in size, and the extended F1500W PSF, including its secondary and tertiary peaks, spreads across a large fraction of the subarray (Fig.~\ref{fig:Background_old_v3_miri_integration_100_aperture_plot}). Most of the encircled energy is contained in the central peak of the PSF, while increasing the aperture to include the outer PSF structure also adds more background and can therefore lower the S/N. In addition, some flux from the extended PSF falls within the background annulus \citep{Gordon2025}. We tested a range of aperture radii after removing the first 30 minutes of the observation to account for the initial time-dependent ramp. As shown in Fig.~\ref{fig:GJ357b_occultation_plots}, the eclipse depth flattens around 5 pixels, where the uncertainty also reaches a minimum. We therefore use an aperture radius of 5 pixels for the fiducial analysis.

\subsection{Global fit for the GJ~357 system}\label{subsec:globalfit}

Published ephemerides for GJ~357~b predict appreciably different occultation times at the epoch of the MIRI observation \citep{Kokori2023,Oddo2023}. Assuming a circular orbit, propagating the reported periods and transit epochs to 11 May 2024 gives predicted occultation mid-times that differ by approximately 15 minutes. An inaccurate eclipse time can affect the recovered depth through its covariance with the instrumental baseline and the placement of ingress and egress. We therefore perform a global fit of the GJ~357 system to obtain a self-consistent ephemeris and set of orbital, planetary, and stellar parameters for use in the eclipse analysis. The orbital geometry determines the predicted eclipse duration and ingress and egress times, while its uncertainty is propagated into the eclipse fit. The brightness temperature does not depend directly on $a/R_\star$ or $i$, but primarily on the measured eclipse depth, $R_{\rm p}/R_\star$, and the adopted stellar spectrum. To improve the constraints on the system parameters, we combine TESS photometry, JWST/NIRSpec transit observations, JWST/MIRI secondary-eclipse observations, radial velocities from CARMENES, HARPS, HIRES, PFS, and UVES, and Gaia parallaxes. The JWST/NIRISS observations of GJ~357~b in transit begin part-way through the transit event \citep{Taylor_2025}; we opt not to include these data in the global fit.

\subsubsection{JWST/NIRSpec transit reduction}

The JWST/NIRSpec transit observations of GJ~357\,b \citep[GO~2512;][]{AdamsRedai2025} were independently reduced for inclusion in the global fit. The reduction follows the same general framework used in the independent NIRSpec analyses of \citet{Chloe_Fisher_2025}, including column-wise 1/f correction, exclusion of the final group during ramp fitting, and pixel-weighted aperture extraction, while the extraction parameters were optimized specifically for the GJ~357\,b observations.

\subsubsection{Global fit methodology}

We perform a global fit including data from TESS Sectors 8, 35, 62, and 89; the reduced JWST/NIRSpec transit observations described above; the JWST/MIRI secondary eclipse presented in this work; radial velocities from CARMENES, HARPS, HIRES, PFS, and UVES \citep{Luque2019}; and parallaxes from Gaia \citep{GaiaDR32023}.

We use the \texttt{EXOFASTv2} code to simultaneously fit the system parameters and stellar spectral energy distribution \citep{Eastman2019}, including priors from empirical relations between $K_\mathrm{s}$-band magnitudes and M-dwarf masses and radii \citep{Mann2015,Mann2019}. There are 63 free parameters in the fit, and we run \texttt{EXOFASTv2} with \texttt{NSTEPS} = 7,500 and \texttt{NTHIN} = 300.

We reach the standard Gelman--Rubin convergence criterion of $\hat{r} < 1.01$ for all but three parameters: (1) the mass of the non-transiting planet GJ~357\,c; (2) the time of conjunction of GJ~357\,c; and (3) the jitter associated with the PFS radial-velocity data, which include observations obtained at both lower ($R\sim80,000$) and higher ($R\sim150,000$) spectral resolution \citep{Luque2019}.

Results from the \texttt{EXOFASTv2} global fit can be found in Table~\ref{tab:globalfit}. The resulting posterior distributions are used as priors in the subsequent analysis of the MIRI secondary eclipse and in the construction of the planetary emission models.

\subsection{Independent reductions and analyses}

Given the unusually deep eclipse measured for GJ~357\,b, together with the presence of detector artefacts near the target PSF and the elevated solar activity during the observation, we performed two fully independent reductions and analyses of the dataset. These analyses use different calibration procedures, photometric extraction methods, and light-curve fitting frameworks. 

\subsubsection{Independent reduction I} \label{IR1}

For the independent analysis, we start the reduction with the \texttt{uncal} files, which were divided into two segments. The files were processed using the \texttt{Eureka!} pipeline version 1.2 \citep{Bell2022}, which wraps the \texttt{jwst} pipeline for Stages 1 and 2 \citep{bushouse_2022_7229890}. We use CRDS context 1364. Stage 1 consists of detector-level calibration, where we correct for the electromagnetic interference (EMI) noise affecting the MIRI data, saturation detection, linearity, dark current, and jump detection step with a threshold of 20$\sigma$. We discarded the first and last groups during ramp fitting. The resulting \texttt{rateints} files were processed in Stage 2, performing flat field correction. The output of Stage 2 are the \texttt{calints} files. In Stage 3, we restrict the subarray region of interest to pixels 16 to 71 in the x-axis and from pixels 1 to 64 in the y-axis. We performed aperture photometry with aperture radii from 3 to 15 pixels, finding that a radius of 5 pixels minimises the median absolute deviation (MAD). An annulus with inner radius of 12 pixels and outer radius of 20 pixels is used for the background, ensuring that the outer annulus radius stays within the subarray region. Finally in Stage 4 we produce the lightcurve and clip outliers above 10$\sigma$ using a rolling median with a width of 10 integrations. 

The lightcurve analysis was performed using an Hamiltonian Monte Carlo routine implemented in the \texttt{PYMC} package \citep{Salvatier2016}. The beginning of the time-series is affected by a steep upward ramp effect known to affect MIRI observations. Rather than masking the start of the visit, we fit the ramp with a single exponential with free parameters describing the timescale, offset, and a multiplicative factor controlling the sign of the ramp. In addition to the exponential function, we fit for the occultation depth following \citet{Meier2025} and a detrending function composed of a linear trend in time and background level. We also tested removing the first 30 minutes of observation and fit only for a linear function, without the exponential term. However, since it is not possible to make a quantitative comparison of models with different numbers of data points using information criteria such as WAIC or Leave-one-out cross-validation \citep{Vehtari2015} for different datasets, we rely on the residual root mean square (RMS). In this case, the model with the complete dataset has a residual RMS of 670 ppm, while the trimmed dataset fitting a linear trend has a residual RMS of 661 ppm, thus we proceed the analysis trimming the ramp. We report an occultation depth of 197 $\pm$ 17 ppm.

\subsubsection{Independent reduction II} \label{IR2}

The input dataset consists of two segments of uncalibrated files (\texttt{uncal}). We performed the detector-level calibration by using the JWST Calibration pipeline v.1.13.4. We ran the emicorr step to subtract the 10.04 Hz EMI noise present in MIRI Imaging data. All subsequent steps (saturation detection, linearity model, dark current) were conducted in a standard fashion. Before proceeding with the ramp fitting, we employed the jump detection step as implemented in \texttt{transitspectroscopy} v.0.4.0 \citep{Espinoza:2022}. We elected to discard the first and last groups during ramp fitting, which often display flux inconsistencies with all-group ramp fit model. We completed the data reduction with JWST Calibration pipeline's stage 2 resulting in high-level calibrated data files (calints) that we use as inputs for the data analysis.
\newline

We first identified bad pixel values that we replaced with values interpolated using flux values from nearest neighbours with a Gaussian weighting. We then computed the star centroid position and performed aperture photometry in a set of apertures ranging from 3 to 20 pixels in radii. We used a background aperture located at 30 pixels from the source with a size of 15 pixels to avoid secondary contribution from the source's PSF. The extracted data all display a strong ramp with a duration of $\sim$30 min that we discarded. We find that the remaining data exhibit a residual downward slope that we fit with a linear trend simultaneously to a planet occultation model in an MCMC framework previously described in \citet[e.g.][]{Gillon2017, Demory2023}. We analyse all sets of time-series (each with flux extracted using different aperture sizes) and find that an aperture radius of 5 pixels yield the smallest red noise contribution, which we find negligible. We measure an occultation depth of 179$_{-16}^{+17}$ ppm for an unbinned RMS of 676 ppm with no noticeable correlated noise.

\subsubsection{Comparing independent data reductions and analyses}

The two independent reductions differ from the primary reduction in their pipeline versions, jump-detection treatment, background estimation, ramp handling, and light-curve fitting frameworks. Nevertheless, both recover occultation depths consistent with the primary measurement within $1\sigma$. We therefore use these reductions as external checks on the robustness of the fiducial detection rather than combining them into a single weighted mean, since all reductions are based on the same underlying observation and are not statistically independent.

\subsection{Eclipse fitting}

The fiducial analysis uses the light curve extracted with the 5-pixel aperture described in Section~\ref{reduction_occultation_fiducial}, excludes the first 30 minutes of data to remove the initial detector-settling ramp, and adopts a linear baseline model. Orbital and ephemeris parameters are allowed to vary under Gaussian priors derived from the posterior distributions of the global \texttt{EXOFASTv2} fit (Section~\ref{subsec:globalfit}). This fiducial analysis yields an occultation depth of $200.5 \pm 12.7$\,ppm.

\subsubsection{High solar activity: jump detection and pixel-level systematics}

For each individual observation of the Hot Rocks Survey targets, we calculated the jump rate as follows. We used the combined uncal files for each observation to calculate the rate of strong detector jumps per pixel over time. To detect sudden jumps in data number, we calculated the difference between each pair of successive groups in the full subarray, excluding the first and last groups, reference pixels, and one pixel along the outer boundary of each subarray. For each pixel, we subtracted the median difference between successive groups to remove the flux accumulated from the PSF. We then counted the number of times a pixel experienced a sudden jump above a given data number threshold and divided the total number of events by the number of included pixels and the observation duration. This gives the frequency of jumps above a given threshold per pixel per hour. Table~\ref{tab:jump_rates} shows, for each Hot Rocks Survey target, the observation with the highest jump rate above 5,000 DN. The jump rate during the GJ~357 observation is substantially higher than for the other survey targets, reaching nearly an order of magnitude higher at the $>5,000$ DN threshold.

\begin{table}
\caption{Jump rates measured for the Hot Rocks targets.}
\label{tab:jump_rates}
\centering
\begin{tabular}{lcccc}
\hline\hline
Target &
\multicolumn{3}{c}{Jump rate (pixel$^{-1}$ h$^{-1}$)} &
PSF rate \\
\cline{2-4}
 & $>5000$ & $>10000$ & $>16000$ & $>5000$ \\
\hline
GJ-3473   & 0.00280 & $4.95\times10^{-4}$ & $1.79\times10^{-4}$ & 0.227 \\
TOI-1468  & 0.00273 & $4.50\times10^{-4}$ & $1.77\times10^{-4}$ & 0.221 \\
LHS-1140  & 0.00279 & $4.99\times10^{-4}$ & $1.99\times10^{-4}$ & 0.226 \\
L-231-32  & 0.00650 & $9.71\times10^{-4}$ & $2.66\times10^{-4}$ & 0.526 \\
LHS-1478  & 0.00297 & $5.36\times10^{-4}$ & $1.88\times10^{-4}$ & 0.240 \\
LTT-3780  & 0.00248 & $4.58\times10^{-4}$ & $1.75\times10^{-4}$ & 0.201 \\
HD-260655 & 0.00194 & $5.74\times10^{-4}$ & $4.30\times10^{-4}$ & 0.157 \\
L-98-59   & 0.00209 & $3.54\times10^{-4}$ & $1.77\times10^{-4}$ & 0.169 \\
\textbf{GJ-357} &
\textbf{0.0565} &
\textbf{0.00557} &
\textbf{0.00124} &
\textbf{4.58} \\
\hline
\end{tabular}
\tablefoot{Rates are given in events per pixel per hour above the indicated data number threshold. For targets with multiple observations, the observation with the highest $>5,000$ jump rate is shown. The final column gives the rate above 5,000 within the PSF region.}
\end{table}

To analyse the impact of the increased solar activity during the observation, we varied the outlier rejection of the \texttt{jump} step using 4$\sigma$, 8$\sigma$, 16$\sigma$, and 32$\sigma$ thresholds. These thresholds are applied to groups within each integration to identify jumps in the detector ramp, rather than to the extracted light curve. Flagged groups are accounted for in the subsequent \texttt{ramp\_fit} step, and the affected integrations are generally retained in the final light curve. We adopt the 4$\sigma$ threshold for the fiducial reduction. We also inspected the individual pixel light curves, motivated by the pixel-level analysis of \citet{Fortune2025}, which shows that systematics affecting individual pixels can remain partly hidden in the aperture-summed light curve.

\begin{figure}[tbp]
\centering
\includegraphics[width=0.49\textwidth]{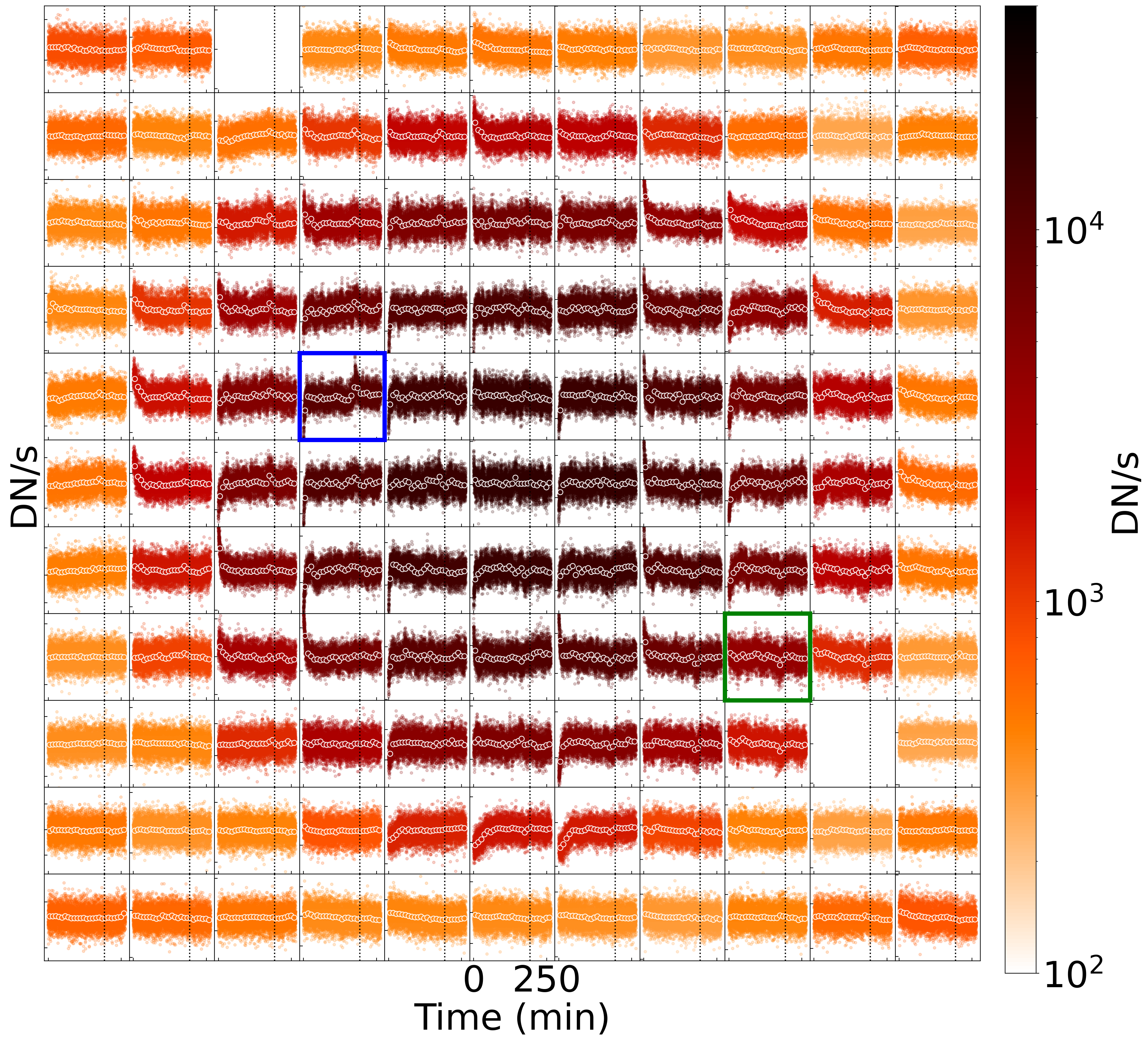}
\caption{Pixel light curves for an 11$\times$11 pixel grid around the stellar PSF, using a 4$\sigma$ jump-rejection threshold after the \textit{ramp\_fit} step. Each mini-plot shows the light curve of an individual pixel over time. The dotted vertical line indicates the transition between segments 001 and 002. The blue and green boxes highlight the pixels containing the most prominent flux excursions. Replacing the affected time samples by interpolation between neighbouring values in the blue pixel, the green pixel, and both pixels simultaneously gives eclipse depths of $205 \pm 13$ ppm, $199 \pm 12$ ppm, and $201 \pm 13$ ppm, respectively, all consistent with the fiducial value of $200.5 \pm 12.7$ ppm.}
\label{fig:Print_Pixel_LightCurves_GJ357b_Obs06}
\end{figure}

Figure~\ref{fig:Print_Pixel_LightCurves_GJ357b_Obs06} shows the digital numbers per second (DN/s) as a function of time for an 11$\times$11 px grid centred on the brightest pixel of the stellar PSF after the \textit{ramp\_fit} step. High-flux pixels (darker reds) show the expected initial detector settling ramp. These ramps can be either increasing or decreasing depending on the position of the pixel relative to the PSF. Similar variations in the strength and the sign of the initial MIRI settling ramp are discussed in \citet{Fortune2025}. We additionally inspected individual pixel light curves and identified two pixels that show prominent flux excursions with opposite signs. These features occur during an observation obtained under elevated solar activity, but their origin cannot be determined from the pixel light curves alone.

To test whether these pixel-level flux excursions affect the measured eclipse depth, we replaced the affected time samples in each highlighted pixel by interpolating between the neighbouring values in that pixel's time series. We performed this test for each pixel separately and for both pixels simultaneously, and then repeated the eclipse fit using the same 30 min cut as in the fiducial analysis. The resulting eclipse depths are consistent with the fiducial value within $1\sigma$, indicating that these two prominent excursions alone do not drive the measured eclipse depth. Despite the structure seen in individual pixel light curves, these features are much less apparent in the summed aperture photometry.

Overall, the eclipse depth remains stable across the tested jump-rejection thresholds, the pixel-level interpolation tests, and the ramp-model choices discussed above. We therefore find no evidence that, for this data set, the measured eclipse depth is driven by the identified pixel-level flux excursions or by localized pixel-level systematics.

\subsubsection{Lightcurve systematics model}

\begin{figure*}[t]
    \centering

    \includegraphics[width=0.95\linewidth]{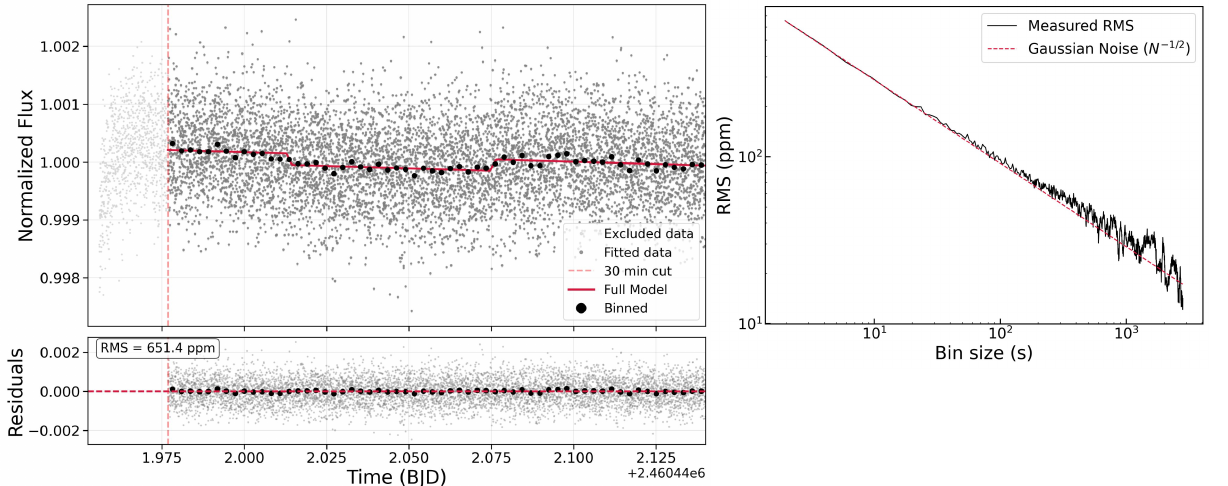}

    \caption{\textit{Left:} Model fit to the \textit{JWST}/MIRI F1500W secondary-eclipse light curve of GJ~357~b. The light gray points show the first 30 minutes of data, which are excluded from the fit due to the initial ramp, while the darker gray points show the data included in the fit. The red dashed line marks the 30-minute cut boundary. The red solid line shows the full model, consisting of the secondary eclipse and a linear baseline trend. The data are binned into 80 equal-width time bins (approximately 2.9 minutes per bin). The lower panel shows the corresponding residuals, with the same time binning applied to the black circles; the unbinned residual RMS is provided. \textit{Right:} Time-averaging diagnostic showing the RMS of the fit residuals as a function of bin size. The black curve shows the measured RMS of the binned residuals, while the red dashed line shows the expected $N^{-1/2}$ scaling for uncorrelated Gaussian noise.}
    \label{fig:gj357b_eclipse_fit}
\end{figure*}

The global fit constrains the orbital ephemeris, system geometry, and occultation depth, but models the MIRI time-series systematics using only a linear trend and offset. Its occultation-depth posterior may therefore not capture the full uncertainty associated with the treatment of these systematics. We therefore test several baseline models to assess the robustness of the inferred eclipse signal: Linear, Linear + GP, Linear + Exp, and Linear + Exp + GP. In all cases, $a/R_\star$, $i$, $P$, $T_0$, $e\cos\omega$, and $e\sin\omega$ are allowed to vary under Gaussian priors derived from the global fit. These parameters are included to propagate uncertainties in the orbital geometry and ephemeris into the eclipse fit rather than to independently constrain them from the single eclipse observation. The eclipse timing provides additional sensitivity primarily to $e\cos\omega$, while $e\sin\omega$ is only weakly constrained by these data.

\begin{figure}[tbp]
    \centering
    \includegraphics[width=0.5\textwidth]{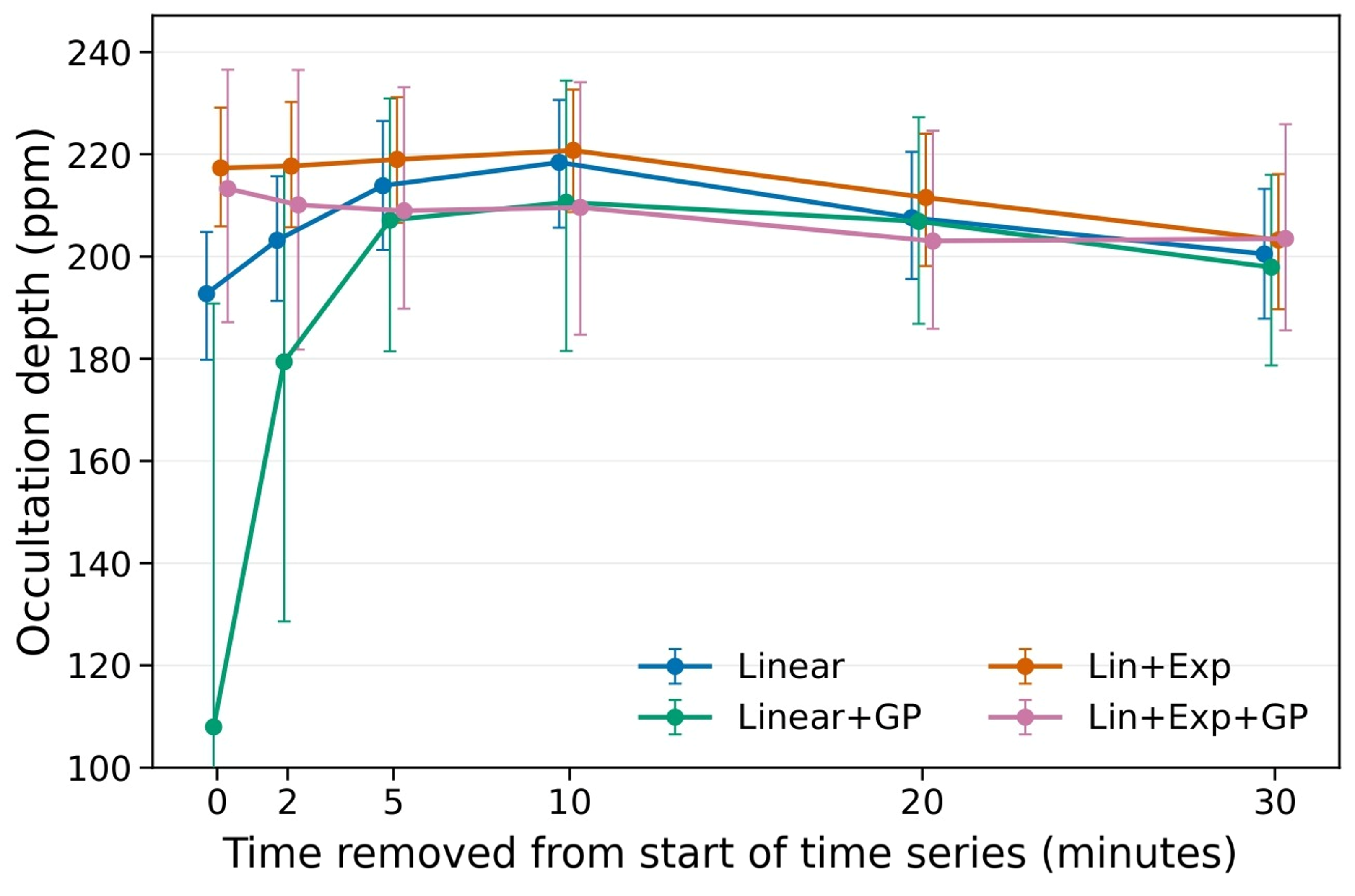}
    \caption{Occultation depth of GJ~357\,b as a function of the initial time removed from the light curve (0--30 min). Colors show the four baseline models, with points offset slightly along the x-axis for clarity. After removing the first 10 min, the inferred occultation depths are consistent across all four models.}
    \label{fig:occ_depth_vs_cutout_AA_jitter}
\end{figure}

Figure~\ref{fig:occ_depth_vs_cutout_AA_jitter} shows the occultation depth as a function of the initial time removed from the light curve (0--30 min). Points show the posterior medians with 16--84\% credible intervals. After removing the first 10 min, all four baseline models give consistent eclipse depths within $1\sigma$.

\setlength{\tabcolsep}{5.5pt}
\renewcommand{\arraystretch}{1.3}
\begin{table}[htbp]
  \centering
  \caption{Comparison of baseline models after removing the first 30 min of the observation. Reported values are posterior medians with 16--84\% credible intervals.}
  \label{tab:occ_depth_30min}
    \begin{tabular}{lccc}
    \hline\hline
    Model & Depth (ppm) & RMS (ppm) & $\Delta$BIC \\
    \hline
    Linear            & $200.5^{+12.7}_{-12.7}$ & 651 & 0.0 \\
    Linear + GP       & $197.9^{+18.1}_{-19.2}$ & 649 & 29.2 \\
    Linear + Exp      & $203.2^{+12.9}_{-13.6}$ & 652 & 32.1 \\
    Linear + Exp + GP & $203.5^{+22.4}_{-17.9}$ & 649 & 66.2 \\
    \hline
    \end{tabular}
\end{table}

Table~\ref{tab:occ_depth_30min} compares the four baseline models after removing the first 30 min of the observation. All models give consistent eclipse depths, ranging from approximately 198 to 204 ppm. Adding an exponential ramp does not improve the RMS, while including a GP reduces the RMS only slightly, from 651 to 649 ppm, and increases the uncertainty on the eclipse depth. The additional model complexity is also strongly penalized by the BIC, with $\Delta$BIC values of 29.2--66.2 relative to the linear model. We therefore adopt the linear baseline model for the fiducial analysis, yielding an occultation depth of $200.5 \pm 12.7$ ppm.

\subsection{Atmospheric and surface forward models}
\label{subsec:atmospheric_models}

To interpret the measured MIRI/F1500W occultation depth, we computed a suite of dayside thermal-emission forward models for GJ~357~b. These models are not intended to exhaust all possible atmospheric or surface states. Instead, they provide a set of physically motivated end-member scenarios against which the measured planet--star contrast can be compared. In particular, the models test whether the observation can be reproduced by a simple airless surface or by standard cloud-free, high-mean-molecular-weight atmospheres expected for a highly irradiated rocky planet after the loss of a primordial H/He envelope.

The atmospheric spectra were calculated with the one-dimensional radiative--convective equilibrium model, \texttt{HELIOS} \citep{Malik2017,Malik2019a,Malik2019b}. The stellar and planetary parameters were fixed to the adopted values from the global fit described in Sections~\ref{subsec:globalfit} and \ref{subsec:stellar_models}, including the stellar radius, planetary radius, surface gravity, orbital separation, and incident stellar spectrum. The same stellar spectrum was used when converting model planet fluxes into planet--star contrasts. The models are cloud-free and assume zero albedo. We omit hazes to keep the comparison focused on a limited set of reference atmospheres. Including hazes would require additional assumptions about their properties and vertical distribution, which the single-band eclipse measurement cannot uniquely constrain. This simplification does not imply that haze effects are negligible, and our conclusions are restricted to the atmospheric models tested here.

Absorption cross sections were computed with \texttt{HELIOS-K} \citep{GrimmHeng2015,Grimm2021} and converted into the opacity tables used by \texttt{HELIOS}. We used CO$_2$ cross sections based on the HITEMP line list \citep{Rothman2010} and H$_2$O cross sections from the BT2 line list \citep{Barber2006}. For the N$_2$+100~ppm CO$_2$ models, N$_2$ acts as the dominant background gas while the CO$_2$ trace abundance provides molecular
infrared opacity. Rayleigh scattering was included for the major background gases following \citet{Cox2000}, \citet{Sneep2005}, and \citet{Thalman2014}.

The atmospheric grid covers high-$\mu$ compositions motivated by rocky-planet outgassing, volatile loss, and photochemistry. We considered pure CO$_2$, pure H$_2$O, and N$_2$ atmospheres containing 100~ppm CO$_2$. We show models at surface pressures of 1, 0.1, 0.01, $10^{-4}$, and $10^{-5}$~cbar. The low-pressure cases probe tenuous, surface-dominated atmospheres that may remain compatible with the featureless transmission spectra of GJ~357~b, while the higher-pressure cases test atmospheres with stronger infrared opacity and more substantial radiative blanketing. Detailed spectra for the model families are shown in Appendix~\ref{app:model_families}.

For comparison with the atmospheric models, we also include an airless-surface calculation. The reference no-atmosphere model represents a zero-albedo, atmosphere-free planet heated only by stellar irradiation. 

Each model emergent spectrum was converted into a wavelength-dependent secondary-eclipse depth using

\begin{equation}
\delta_\lambda =
\left(\frac{R_{\rm p}}{R_\star}\right)^2
\frac{F_{{\rm p},\lambda}}{F_{\star,\lambda}},
\label{eq:monochromatic_eclipse_depth}
\end{equation}

where $F_{{\rm p},\lambda}$ and $F_{\star,\lambda}$ are the planet and stellar surface fluxes per unit wavelength. For comparison with the JWST/MIRI observation, we calculated the band-integrated eclipse depth as

\begin{equation}
\delta_{\rm F1500W} =
\frac{
\int \lambda\,T_{\rm F1500W}(\lambda)\,
F_{\star,\lambda}\,\delta_\lambda\,{\rm d}\lambda
}{
\int \lambda\,T_{\rm F1500W}(\lambda)\,
F_{\star,\lambda}\,{\rm d}\lambda
},
\label{eq:f1500w_band_integrated_depth}
\end{equation}
where $T_{\rm F1500W}(\lambda)$ is the dimensionless F1500W throughput. The factor of $\lambda$ accounts for photon-counting weighting when the spectra are expressed as energy flux densities per unit wavelength; the common factor $1/(hc)$ cancels in the ratio. Figure~\ref{fig:gj357b_atmospheric_models} shows the model spectra together with their corresponding F1500W band-integrated eclipse depths.

\subsection{Stellar spectrum and its impact on the inferred planetary emission}
\label{subsec:stellar_models}

To assess how uncertainties in the model stellar spectrum affect the physical interpretation of the measured eclipse depth, we compare the adopted stellar model to the flux-calibrated stellar spectrum of GJ~357 measured with JWST/NIRISS SOSS and NIRSpec/G395H as part of transit observations \citep[GO~1201, PI Lafreniere; GO~2512 PI Batalha, respectively;][]{Taylor_2025,AdamsRedai2025}. We additionally determine the absolute flux in the MIRI/F1500W bandpass analysed in this work (GO~3730) using the instructions provided in \citet{Gordon2025}. 

\begin{figure*}[t]
    \centering
    \includegraphics[width=\textwidth]{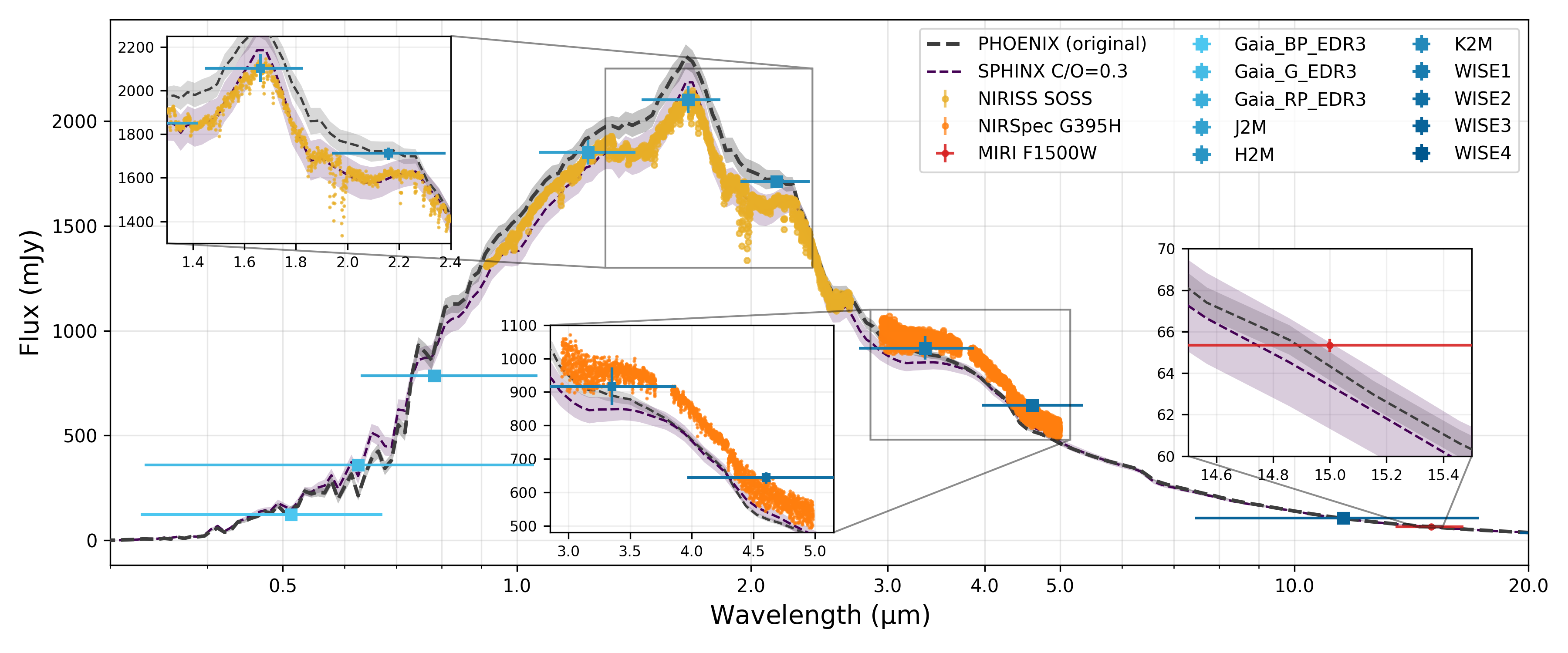}
    \caption{Flux-calibrated JWST data from NIRISS SOSS (GO 1201), NIRSpec G395H (GO 2512), and MIRI F1500W (GO 3730). Data are compared to photometric points, as well as two sets of stellar models (PHOENIX BT-Settl and SPHINX).}
    \label{fig:gj357_stellar_spectrum}
\end{figure*}

We compare the extracted stellar flux from JWST data to the PHOENIX BT-Settl stellar spectrum used to produce the forward models shown in Figure~\ref{fig:gj357b_atmospheric_models} \citep{Husser2013}. The BT-Settl models took in the best-fit stellar parameters and uncertainties from the global fit (Section~\ref{subsec:globalfit}), which included a JWST NIRSpec transit of GJ 357 b to constrain the planet's orbital parameters, but did not include any JWST flux-calibrated data for the star itself to constrain the SED, only measurements from Gaia DR3, 2MASS, and WISE. We find that there is slight disagreement between the BT-Settl model and the flux-calibrated stellar data (see grey-dashed line vs yellow/orange/red data points in Figure~\ref{fig:gj357_stellar_spectrum}). 

For comparison, we produce a set of SPHINX models with a range of C/O ratios \citep{Iyer2023,Iyer2026}. The model SPHINX spectrum with C/O=0.3 and propagated uncertainties matches the flux-calibrated data better in the NIRISS wavelengths, but not the NIRSpec data (Figure~\ref{fig:gj357_stellar_spectrum}). The most significant disagreement is between the 2MASS $K_\mathrm{s}$ photometry and the NIRISS data (see top-left inset in Figure~\ref{fig:gj357_stellar_spectrum}), though this could be due to a bias in the zero-points needed to convert the K$_s$ magnitude of an M dwarf into a flux measurement. We explore a wider range of C/O ratios in the SPHINX models (Appendix~\ref{app:sphinx}) but none are able to exactly match the flux-calibrated data at all wavelengths. It is uncertain whether the disagreement between data and models comes from model uncertainty, missing opacities, calibration uncertainty, or even stellar variability; likely it is some combination of effects.

We explore whether the difference in the original BT-Settl stellar model and the measured stellar flux has a noticeable impact on the derived planetary parameters for GJ 357 b. \texttt{EXOFASTv2} does not allow for SED fitting against a full spectrum, so we convert the flux calibrated spectra from NIRISS and NIRSpec into synthetic photometry based on JWST filter bandpasses in similar wavelength ranges using zero-points from the Spanish Virtual Observatory Filter Profile Service\footnote{\href{https://svo2.cab.inta-csic.es/theory/fps3/}{https://svo2.cab.inta-csic.es/theory/fps3/}} \citep{SVO2012,Rodrigo2020}. We then re-run the \texttt{EXOFASTv2} fit using the JWST synthetic photometric points, as well as the Gaia DR3 photometry to anchor the blue end of the SED. The resulting stellar parameters shift by up to 1--2$\upsigma$, but when all uncertainties are propagated this translates to changes in the planetary properties by less than 1$\upsigma$.

Finally, we assess the effect of adopting the SPHINX stellar spectrum with C/O=0.3 for the no-atmosphere case. For the zero-albedo model, the predicted MIRI/F1500W eclipse depth remains significantly below the measured value.

\section{Results}\label{sec:results}

The analyses described in Section~\ref{sec:datareduction} yield a robust occultation depth of $200 \pm 13$\,ppm in the JWST/MIRI F1500W bandpass. Independent reductions recover consistent depths within $1\sigma$, supporting the robustness of the measurement. We therefore adopt the fiducial value of $200 \pm 13$\,ppm for the following interpretation.

\subsection{Brightness temperature and comparison to forward models}

We measure an F1500W eclipse depth of $200.5 \pm 12.7\,\mathrm{ppm}$, corresponding to a dayside brightness temperature at $15\,\mu\mathrm{m}$ of $T_{\mathrm{b},15\,\mu\mathrm{m}}=923^{+39}_{-38}\,\mathrm{K}$. For comparison, assuming zero Bond albedo and no heat redistribution \citep[$f=2/3$;][]{Cowan2011b,Seager2010},

\begin{equation}
T_{\mathrm{day},0} =
T_{\rm eff,*}
\left(\frac{R_\star}{a}\right)^{1/2}
\left(\frac{2}{3}\right)^{1/4}.
\label{eq:zero_albedo_dayside_temperature}
\end{equation}

For GJ~357~b, this gives $T_{\mathrm{day},0}=686 \pm 14\,\mathrm{K}$.
 The forward models used for this comparison are described in Section~\ref{subsec:atmospheric_models}. Here we compare the measured F1500W eclipse depth to the model spectra and their corresponding band-integrated predictions.

\begin{figure*}[t]
    \centering
    \includegraphics[width=\textwidth]{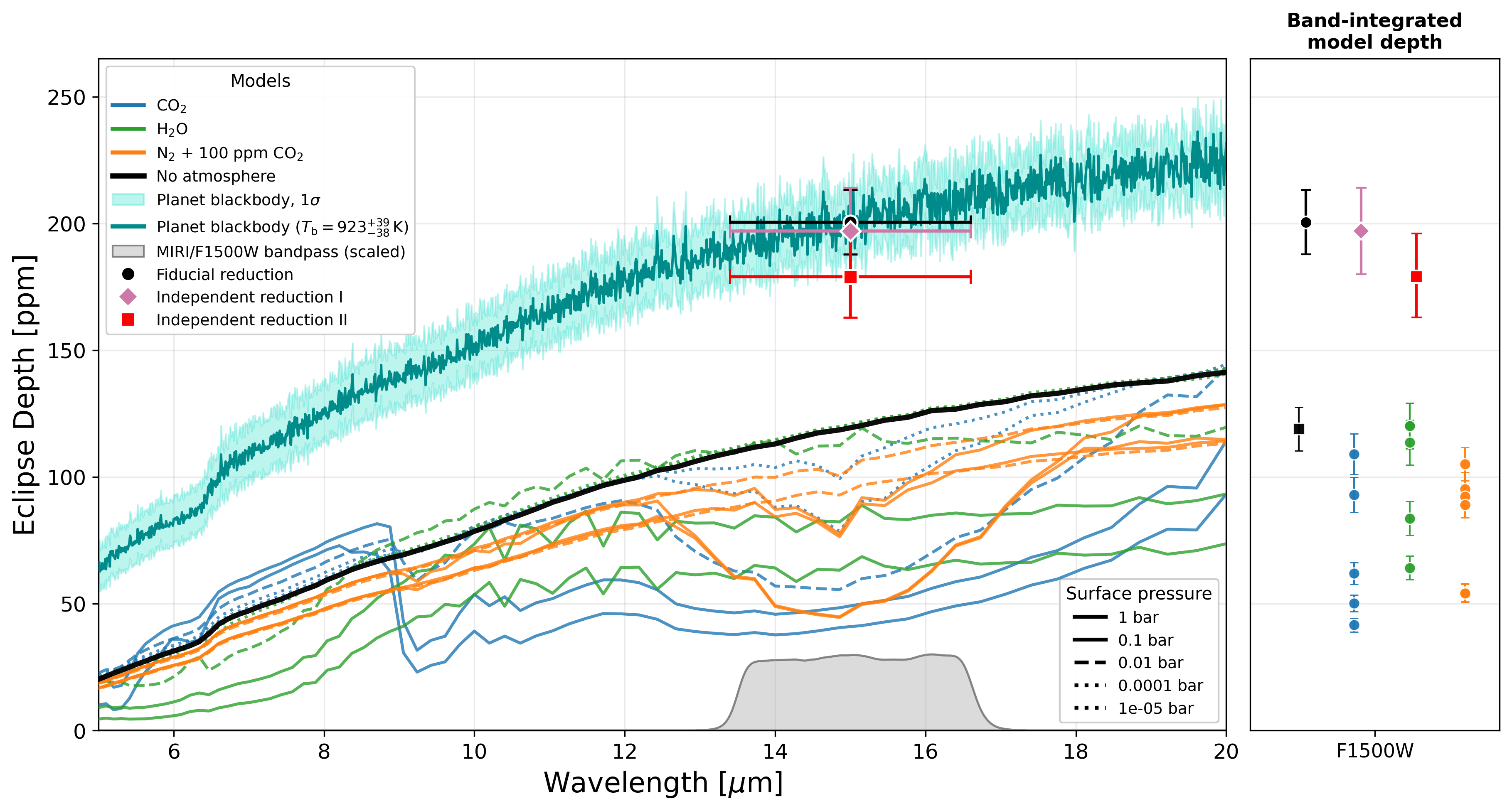}
    \caption{
    Comparison between the measured MIRI/F1500W occultation depth of GJ~357~b and atmospheric forward models.
    The left panel shows model emission spectra for pure CO$_2$, pure H$_2$O, and N$_2$ with 100~ppm CO$_2$, together with the zero-albedo no-atmosphere model (black).
    Line styles distinguish the atmospheric surface pressures of 1, 0.1, 0.01, $10^{-4}$, and $10^{-5}$~bar.
    The shaded region and solid line show the 1$\sigma$ posterior-predictive interval and median, respectively, obtained by modelling the planet as a blackbody, corresponding to a brightness temperature of $T_{\mathrm{b},15\,\mu\mathrm{m}}=923^{+39}_{-38}\,\mathrm{K}$.
    The scaled MIRI/F1500W bandpass is shown in grey for reference.
    The black circle denotes our fiducial reduction ($200.5\pm12.7$~ppm), while the purple diamond and red square show Independent reductions I ($197\pm17$~ppm) and II ($179^{+17}_{-16}$~ppm), respectively.
    Horizontal error bars indicate the approximate wavelength coverage of the F1500W bandpass.
    The right panel shows the corresponding F1500W band-integrated model eclipse depths, calculated using the adopted stellar spectrum and the F1500W photon-counting throughput.
    For the CO$_2$ and H$_2$O models, vertical error bars are obtained by integrating the corresponding lower and upper model bounds over the bandpass; because the N$_2$+100~ppm CO$_2$ grid does not provide corresponding bounds, its uncertainties are estimated from the non-grid models at the same pressure.
    The three independently derived occultation depths are shown alongside the band-integrated model predictions for comparison.
    }
    \label{fig:gj357b_atmospheric_models}
\end{figure*}

Figure~\ref{fig:gj357b_atmospheric_models} compares the measured eclipse depth to the model predictions. The shaded region and solid line show the 1$\sigma$ posterior-predictive interval and median, respectively, obtained by modelling the planet as a blackbody using the adopted stellar spectrum. This corresponds to a brightness temperature of $T_{\mathrm{b},15\,\mu\mathrm{m}}=923^{+39}_{-38}\,\mathrm{K}$ and represents the planetary emission implied by the fiducial measurement, rather than an irradiation-based prediction. In contrast, the thick black curve shows the zero-albedo, no-atmosphere model, which predicts substantially less emission in the F1500W band.

We ran radiative-convective models with \texttt{HELIOS} \citep{Malik2017,Malik2019a,Malik2019b} for high-$\mu$ atmospheres consisting of pure CO$_2$, pure H$_2$O, and N$_2$ with 100~ppm CO$_2$. These compositions provide representative high-$\mu$ atmospheric cases motivated by rocky-planet outgassing and volatile evolution \citep{doi:10.1126/science.259.5097.915,annurev:/content/journals/10.1146/annurev-astro-052920-125632}. We show surface pressures of 1, 0.1, 0.01, $10^{-4}$, and $10^{-5}$\,bar (Appendix~\ref{app:model_families}). Across the model grid, the F1500W-integrated eclipse depths remain below the measured fiducial value. The closest model is the $10^{-4}$\,bar H$_2$O atmosphere, which differs from the fiducial measurement by $5.1\sigma$ when the observational and model uncertainties are combined in quadrature. The same model remains discrepant by $3.9\sigma$ and $3.2\sigma$ for Independent reductions I and II, respectively. The zero-albedo no-atmosphere model differs from the fiducial measurement by $5.2\sigma$.

\section{Discussion}\label{sec:discussion}

The MIRI/F1500W eclipse depth of $200.5 \pm 12.7\,\mathrm{ppm}$ corresponds to a dayside brightness temperature at $15\,\mu\mathrm{m}$ of $T_{\mathrm{b},15\,\mu\mathrm{m}}=923^{+39}_{-38}\,\mathrm{K}$, substantially above the zero-albedo, no-redistribution reference temperature of $T_{\mathrm{day},0}=686 \pm 14\,\mathrm{K}$ \citep{Cowan2011b,Seager2010}. Below we outline plausible explanations and the observations that can distinguish them.

\subsection{Why is the dayside emission so high?}

A single-band brightness temperature does not uniquely map to a physical surface temperature because the emergent spectrum can be strongly non-grey. 

\paragraph{Non-grey surface emission.} 

One possibility is that the dayside emission is shaped by wavelength-dependent surface emissivity. Laboratory measurements and Solar System observations show that silicate surfaces and lava-bearing terrains exhibit pronounced mid-infrared spectral structure, including deviations from blackbody emission around $8$--$15\,\mu\mathrm{m}$ \citep[e.g.][]{Fortin_2024,KUMARI2024115976}. The Christiansen feature at $\sim 7$--$9\,\mu\mathrm{m}$ marks an emissivity maximum, while Reststrahlen bands at $\sim 9$--$12\,\mu\mathrm{m}$ create an emissivity minimum \citep[e.g.][]{henderson2005structure,ramsey1998mineral}. At longer wavelengths, emissivity recovers toward $15$--$20\,\mu\mathrm{m}$. In addition, subsurface temperature gradients in an airless regolith can modify the emergent spectrum through solid-state greenhouse or anti-greenhouse effects, potentially increasing the secondary-eclipse depth above the corresponding blackbody prediction at some wavelengths \citep{lyu2025impactsubsurfacetemperaturegradients}. Such effects could enhance the F1500W flux relative to a grey-surface prediction without requiring every surface region or emitting layer to be hotter than expected from stellar irradiation. Surface roughness may also produce thermal-infrared beaming, preferentially directing thermal emission back toward the star and therefore toward the observer near secondary eclipse \citep{SPENCER1990,Rozitis2011}.

\paragraph{Atmospheric emission from unmodelled compositions.}
A second possibility is that the planet retains a tenuous atmosphere with a composition or thermal structure not represented in the present model grid. Additional opacity alone would not necessarily increase the F1500W emission, since molecular absorption generally shifts the photosphere to cooler atmospheric layers. Enhanced emission in this band could instead arise if the atmosphere develops a temperature inversion, producing molecular bands in emission, or if the wavelength-dependent opacity provides a window through which hotter, deeper layers are observed. The pure CO$_2$, pure H$_2$O, and N$_2$+100~ppm CO$_2$ models considered here do not reproduce the measured F1500W eclipse depth, but broader wavelength coverage is required to test more complex atmospheric structures.

\paragraph{Additional heating and localized hot regions.}
Another possibility is that additional internal heating produces regions of the dayside that are hotter than expected from stellar irradiation alone. Volcanic or magma-bearing surfaces can produce strongly non-uniform thermal emission, as observed for Solar System bodies and predicted for rocky exoplanets \citep[e.g.][]{de_Kleer_2019,DAVIES2015,Kite2016,Hammond_2017}. However, modelling the surface temperature distribution or emitting area required to reproduce the measured F1500W eclipse depth is out of the scope of this work, and we therefore cannot determine whether localized hot regions provide a viable explanation for the measured eclipse depth of GJ~357\,b.

Another candidate for additional heating is electromagnetic induction, which we find is unlikely to provide sufficient additional thermal flux to explain this system \citep{Peng_2025}. The modest eccentricity inferred from the global fit, $e = 0.047^{+0.032}_{-0.030}$, changes the instantaneous stellar irradiation only slightly and cannot account for the observed brightness-temperature excess. It may also generate internal heating through tidal dissipation, although the approximately quadratic dependence on eccentricity makes this a less likely explanation for the full excess. The magnitude of the heating also depends strongly on the planet's interior rheology and tidal dissipation efficiency, and therefore cannot be determined from the eccentricity alone \citep{Jackson2008,DriscollBarnes2015}.

\subsection{Observations needed to break degeneracies}

A broader eclipse spectrum across the mid-infrared is the most direct way to distinguish between a grey, blackbody-like emitter and a structured spectrum caused by atmospheric opacity or surface emissivity. Low-resolution spectroscopy with MIRI/LRS (5--14\,$\mu$m) would be sensitive to silicate-related spectral structure---the Christiansen feature and Reststrahlen bands---and would test whether the high planet--star contrast persists across the band \citep{Rieke2015,Rigby2023}. The spectral shape and any resolved features would help test surface-emissivity and atmospheric-opacity models, although a smooth spectrum alone would not distinguish between them. Such observations would also directly probe deviations from grey emission expected for bare rock surfaces or thin atmospheres.

Though a larger investment of telescope time, phase-resolved photometry would test heat transport. A synchronously rotating bare planet with inefficient heat redistribution is expected to exhibit a large day--night thermal contrast and therefore a strongly varying thermal phase curve, whereas an atmosphere capable of efficient heat transport would reduce the day--night contrast and flatten the phase variation \citep[e.g.][]{Kreidberg2018,Hammond2025}.

Repeating eclipse observations would further establish whether the measured depth is stable over time, helping to distinguish persistent physical emission from visit-specific systematics or transient Solar, stellar, and instrumental effects.


\section{Conclusion}\label{sec:conclusion}

As part of the Hot Rocks Survey (GO~3730), we obtained one JWST/MIRI F1500W secondary-eclipse time series of GJ~357~b, a rocky exoplanet orbiting a nearby M dwarf. We measure an occultation depth of $200.5 \pm 12.7\,\mathrm{ppm}$, corresponding to a dayside brightness temperature in the MIRI/F1500W band of $T_{\mathrm{b},15\,\mu\mathrm{m}}=923^{+39}_{-38}\,\mathrm{K}$. For comparison, the zero-Bond-albedo, no-redistribution approximation gives a disk-integrated bolometric dayside reference temperature of $686 \pm 14\,\mathrm{K}$. The measured F1500W eclipse depth exceeds the prediction of the tested zero-albedo, no-atmosphere model. The result is robust to different choices in the data reduction and analysis, including alternative treatments of the initial detector ramp, and is confirmed by two independent reductions of the same dataset.

We also performed a new global analysis of the GJ~357 system using \texttt{EXOFASTv2} \citep{Eastman2019}, incorporating \emph{TESS} photometry, a JWST/NIRSpec transit, radial velocities from CARMENES, HARPS, HIRES, PFS, and VLT/UVES, and the \emph{Gaia} parallax. The resulting stellar and planetary parameters were used consistently in the eclipse analysis and forward models. Newly extracted stellar spectra from NIRISS and NIRSpec indicate some disagreement with the stellar spectral models used in the forward calculations, but the corresponding changes do not remove the discrepancy between the measured eclipse depth and the predicted no-atmosphere case.

The measured F1500W depth is larger than predicted by all of the atmospheric models considered here. The closest model, a $10^{-4}$\,bar H$_2$O atmosphere, differs from the fiducial measurement by $5.1\sigma$ and remains discrepant at $3.2\sigma$ even for Independent reduction II. The zero-albedo no-atmosphere model differs from the fiducial measurement by $5.2\sigma$. Possible explanations include non-grey surface emission, including subsurface temperature-gradient effects, atmospheric opacity sources not represented in the present model grid, or additional internal heating and spatially non-uniform dayside temperatures.

In the broader context of thermal-emission measurements of rocky planets around M dwarfs, GJ~357\,b is unusual. Several previous measurements have disfavoured substantial atmospheres and are broadly consistent with inefficient heat redistribution from an airless or nearly airless surface \citep[e.g.][]{Kreidberg2018,Crossfield2022,Greene2023,Zieba2023,Fortune2025,Allen2025,Holmberg2026}. At the same time, a smaller number of systems, including TOI-1468\,b, show dayside emission that is higher than expected from standard atmospheric models \citep{Meier2025}. Unlike several other targets that are broadly consistent with a bare-rock interpretation, the measured F1500W emission of GJ~357\,b is also not reproduced by the simple no-atmosphere model considered here. GJ~357\,b therefore remains difficult to explain with either of these simple scenarios. Additional mid-infrared eclipse observations, particularly with broader wavelength coverage, would help determine whether the excess emission is a broad thermal effect or instead has a wavelength-dependent origin related to the surface or atmosphere.

\section*{Data availability}

The raw JWST observations used in this work are publicly available
from the Mikulski Archive for Space Telescopes under program
GO~3730. The TESS observations are also publicly available from
the Mikulski Archive for Space Telescopes. The reduced data
products, posterior samples, and model outputs underlying this
article are available from the corresponding author upon reasonable
request and will be deposited in a public repository during the
peer-review process.

\begin{acknowledgements}
This work is based on observations with the NASA/ESA/CSA James Webb Space Telescope obtained at the Space Telescope Science Institute, which is operated by the Association of Universities for Research in Astronomy, Incorporated, under NASA contract NAS5-03127. Support for program number 3730 was provided through a grant from the STScI under NASA contract NAS5-03127.
We thank Aarynn Carter for helpful conversations about the NIRISS/SOSS pipeline. This research has made use of the SVO Filter Profile Service ``Carlos Rodrigo'', funded by MCIN/AEI/10.13039/501100011033/ through grant PID2023-146210NB-I00.

B.-O. D. acknowledges support from the Swiss State Secretariat for
Education, Research and Innovation (SERI) under contract number
MB22.00046.

JMM acknowledges support from the Horizon Europe Guarantee Fund,
grant EP/Z00330X/1.

EMV acknowledges financial support from the Swiss National Science
Foundation (SNSF) Mobility Fellowship under grant no.
P500PT\_225456/1.

NHA acknowledges support by the National Science Foundation Graduate
Research Fellowship under Grant No. DGE1746891.

PCA acknowledges support from the Carlsberg Foundation,
grant CF22-1254.

A.G. acknowledges support from the Trottier Family Foundation through the Trottier Postdoctoral Fellowship at the Institute for Research on Exoplanets (IREx)

NPG gratefully acknowledges support from Science Foundation Ireland
and the Royal Society through a University Research Fellowship
(URF\textbackslash R\textbackslash 201032).

KH acknowledges partial financial support from the European Research
Council (ERC) Geoastronomy Synergy Grant (grant number 101166936).

\end{acknowledgements}

\bibliographystyle{aa} 
\bibliography{bibliography} 

\appendix
\nolinenumbers

\onecolumn
\clearpage

\section{Global fit posterior distributions}
\label{appendix:globalfit}

Here we include the full set of posterior distributions from the global \texttt{EXOFASTv2} fit of the GJ 357 system, including TESS photometry, NIRSpec transit data, the MIRI secondary eclipse that is the focus of this work, and RVs. More information about this fit can be found in Section~\ref{subsec:globalfit}. Note that GJ 357 c and d are not found to transit.

\setstretch{1.0}
\setlength{\tabcolsep}{0.5pt}
\begin{longtable}{l@{\hskip 1pt}l@{\hskip 4pt}c@{\hskip 4pt}c@{\hskip 4pt}c@{\hskip 4pt}}
\caption{\texttt{EXOFASTv2} median values and 68\% confidence interval for the GJ 357 system}
\label{tab:globalfit} \\
\hline\hline
Parameter & Description & \multicolumn{3}{c}{Values} \\
\hline
\vspace{5mm}
\endfirsthead 
\caption{continued.}\\ 
\hline\hline 
Parameter & Description & \multicolumn{3}{c}{Values} \\
\hline
\endhead
\hline
\endfoot
Stellar Parameters:& & & GJ 357 & \smallskip\\
~~~~$M_*$\dotfill & \multicolumn{2}{c}{Mass ($M_\odot$)\dotfill }&$0.3455^{+0.0078}_{-0.0077}$\\
~~~~$R_*$\dotfill & \multicolumn{2}{c}{Radius (\rsun)\dotfill }&$0.3382^{+0.0093}_{-0.0087}$\\
~~~~$R_{*,SED}$\dotfill &\multicolumn{2}{c}{Radius$^{1}$ (\rsun)\dotfill }&$0.3713^{+0.0068}_{-0.0087}$\\
~~~~$L_*$\dotfill &\multicolumn{2}{c}{Luminosity (\lsun)\dotfill }&$0.01621^{+0.00056}_{-0.00055}$\\
~~~~$F_{Bol}$\dotfill &\multicolumn{2}{c}{Bolometric Flux (cgs)\dotfill}&$5.83\pm0.20 \times 10^{-9}$\\
~~~~$\rho_*$\dotfill &\multicolumn{2}{c}{Density (cgs)\dotfill}&$12.6\pm1.0$\\
~~~~$\log{g}$\dotfill &\multicolumn{2}{c}{Surface gravity (cgs)\dotfill}&$4.918^{+0.024}_{-0.025}$\\
~~~~$T_{\rm eff}$\dotfill &\multicolumn{2}{c}{Effective temperature (K)\dotfill }&$3540\pm51$\\
~~~~$T_{\rm eff,SED}$\dotfill &\multicolumn{2}{c}{Effective temperature$^{1}$ (K)\dotfill}&$3396^{+40}_{-35}$\\
~~~~$[{\rm Fe/H}]$\dotfill &\multicolumn{2}{c}{Metallicity (dex)\dotfill }&$-0.50^{+0.22}_{-0.21}$\\
~~~~$K_S$\dotfill &\multicolumn{2}{c}{Absolute Ks-band mag (mag)\dotfill }&$6.598\pm0.020$\\
~~~~$k_S$\dotfill &\multicolumn{2}{c}{Apparent Ks-band mag (mag)\dotfill }&$6.471\pm0.020$\\
~~~~$A_V$\dotfill &\multicolumn{2}{c}{V-band extinction (mag)\dotfill }&$0.098^{+0.072}_{-0.067}$\\
~~~~$\sigma_{SED}$\dotfill &\multicolumn{2}{c}{SED photometry error scaling \dotfill }&$1.65^{+0.74}_{-0.44}$\\
~~~~$\varpi$\dotfill &\multicolumn{2}{c}{Parallax (mas)\dotfill}&$106.015\pm0.025$\\
~~~~$d$\dotfill &\multicolumn{2}{c}{Distance (pc)\dotfill}&$9.4326\pm0.0022$\\
\smallskip\\\multicolumn{2}{l}{Planetary Parameters:}&b&c&d\smallskip\\
~~~~$P$\dotfill &Period (days)\dotfill &$3.93060494^{+0.00000062}_{-0.00000058}$&$9.1253^{+0.0015}_{-0.0013}$&$55.606^{+0.062}_{-0.053}$\\
~~~~$R_P$\dotfill &Radius (\re)\dotfill &$1.152^{+0.034}_{-0.032}$&--&--\\
~~~~$M_P$\dotfill &Mass (\me)\dotfill &$2.05\pm0.43$&--&--\\
~~~~$T_C$\dotfill &Observed Time of conjunction$^{2}$ ($\mathrm{BJD}_{\mathrm{TDB}}$)\dotfill &$2459272.67580^{+0.00015}_{-0.00016}$&$2458314.56^{+0.75}_{-0.62}$&$2458320.0^{+5.7}_{-4.6}$\\
~~~~$T_C$\dotfill &Model Time of conjunction$^{2,3}$ (\tjdtdb)\dotfill &$2459272.67561^{+0.00015}_{-0.00016}$&$2458314.56^{+0.75}_{-0.62}$&$2458320.0^{+5.7}_{-4.6}$\\
~~~~$T_T$\dotfill &Model time of min proj sep$^{3,4,5}$ (\tjdtdb)\dotfill &$2460196.367775\pm0.000058$&--&--\\
~~~~$T_0$\dotfill &Obs time of min proj sep$^{4,6,7}$ ($\mathrm{BJD}_{\mathrm{TDB}}$)\dotfill &$2460196.367963\pm0.000058$&$-6232.55^{+0.92}_{-1.1}$&$-7117.6^{+6.8}_{-8.0}$\\
~~~~$a$\dotfill &Semi-major axis (AU)\dotfill &$0.03420^{+0.00025}_{-0.00026}$&$0.05997\pm0.00045$&$0.2001\pm0.0015$\\
~~~~$i$\dotfill &Inclination (Degrees)\dotfill &$89.11^{+0.33}_{-0.22}$&--&--\\
~~~~$e$\dotfill &Eccentricity \dotfill &$0.047^{+0.032}_{-0.030}$&$0.115^{+0.12}_{-0.081}$&$0.18^{+0.17}_{-0.12}$\\
~~~~$\omega_*$\dotfill &Arg of periastron (Degrees)\dotfill &$81.3^{+4.1}_{-18}$&$-150^{+100}_{-110}$&$7^{+62}_{-63}$\\
~~~~$\dot{\omega}_{\rm GR}$\dotfill &Computed GR precession ($^\circ$/century)\dotfill &$1.004\pm0.015$&$0.2502^{+0.011}_{-0.0055}$&$0.01252^{+0.0012}_{-0.00042}$\\
~~~~$T_{\rm eq}$\dotfill &Equilibrium temp$^{8}$ (K)\dotfill &$537.0^{+5.0}_{-4.9}$&$405.6^{+3.8}_{-3.7}$&$222.0^{+2.1}_{-2.0}$\\
~~~~$K$\dotfill &RV semi-amplitude (m/s)\dotfill &$1.69\pm0.35$&$2.21^{+0.35}_{-0.36}$&$2.36^{+0.42}_{-0.38}$\\
~~~~$R_P/R_*$\dotfill &Radius of planet in stellar radii \dotfill &$0.03123\pm0.00018$&--&--\\
~~~~$a/R_*$\dotfill &Semi-major axis in stellar radii \dotfill &$21.75^{+0.59}_{-0.60}$&$38.1\pm1.0$&$127.2^{+3.4}_{-3.5}$\\
~~~~$\delta$\dotfill &$\left(R_P/R_*\right)^2$ \dotfill &$0.000975\pm0.000011$&--&--\\
~~~~$\tau$\dotfill &In/egress transit duration (days)\dotfill &$0.00181^{+0.00013}_{-0.00012}$&--&--\\
~~~~$T_{14}$\dotfill &Total transit duration (days)\dotfill &$0.05380^{+0.00018}_{-0.00017}$&--&--\\
~~~~$T_{FWHM}$\dotfill &FWHM transit duration (days)\dotfill &$0.05198\pm0.00013$&--&--\\
~~~~$b$\dotfill &Transit impact parameter \dotfill &$0.321^{+0.079}_{-0.12}$&--&--\\
~~~~$b_S$\dotfill &Eclipse impact parameter \dotfill &$0.352^{+0.074}_{-0.12}$&--&--\\
~~~~$\tau_S$\dotfill &In/egress eclipse duration (days)\dotfill &$0.00201^{+0.00014}_{-0.00012}$&--&--\\
~~~~$T_{S,14}$\dotfill &Total eclipse duration (days)\dotfill &$0.0583^{+0.0038}_{-0.0030}$&--&--\\
~~~~$T_{S,FWHM}$\dotfill &FWHM eclipse duration (days)\dotfill &$0.0563^{+0.0038}_{-0.0030}$&--&--\\
~~~~$\rho_P$\dotfill &Density (cgs)\dotfill &$7.3^{+1.7}_{-1.6}$&--&--\\
~~~~$logg_P$\dotfill &Surface gravity (cgs)\dotfill &$3.179^{+0.086}_{-0.10}$&--&--\\
~~~~$\Theta$\dotfill &Safronov Number \dotfill &$0.0124\pm0.0026$&--&--\\
~~~~$\langle F\rangle$\dotfill &
Incident Flux ($10^9\,\mathrm{erg\,s^{-1}\,cm^{-2}}$)\dotfill &$0.01881^{+0.00071}_{-0.00068}$&$0.00600^{+0.00027}_{-0.00030}$&$0.000529^{+0.000028}_{-0.000043}$\\
~~~~$T_S$\dotfill &Observed Time of eclipse$^{2}$ ($\mathrm{BJD}_{\mathrm{TDB}}$)\dotfill &$2459270.7279^{+0.0014}_{-0.0018}$&$2458319.02^{+0.67}_{-0.65}$&$2458297.1^{+3.9}_{-3.6}$\\
~~~~$T_S$\dotfill &Model Time of eclipse$^{2,3}$ (\tjdtdb)\dotfill &$2459270.7281^{+0.0014}_{-0.0018}$&$2458319.02^{+0.67}_{-0.65}$&$2458297.1^{+3.9}_{-3.6}$\\
~~~~$T_E$\dotfill &Model time of sec min proj sep$^{3,4,5}$ (\tjdtdb)\dotfill &$2460756.4968^{+0.0014}_{-0.0018}$&--&--\\
~~~~$T_{E,0}$\dotfill &Obs time of sec min proj sep$^{4,6,7}$ ($\mathrm{BJD}_{\mathrm{TDB}}$)\dotfill &$2460756.4966^{+0.0014}_{-0.0019}$&$2445.57^{+0.42}_{-0.36}$&$2502.3^{+2.8}_{-2.4}$\\
~~~~$e\cos{\omega_*}$\dotfill & \dotfill &$0.00712^{+0.00056}_{-0.00075}$&$-0.013^{+0.086}_{-0.12}$&$0.11^{+0.20}_{-0.12}$\\
~~~~$e\sin{\omega_*}$\dotfill & \dotfill &$0.046^{+0.033}_{-0.031}$&$-0.011^{+0.085}_{-0.12}$&$0.012^{+0.12}_{-0.098}$\\
~~~~$M_P\sin i$\dotfill &Minimum mass (\me)\dotfill &$2.05\pm0.43$&$3.51\pm0.57$&$6.7\pm1.1$\\
~~~~$M_P/M_*$\dotfill &Mass ratio \dotfill &$1.78\pm0.37 \times 10^{-5}$&$3.57^{+0.57}_{-0.58} \times 10^{-5}$&$0.000106\pm0.000017$\\
~~~~$d/R_*$\dotfill &Separation at mid transit \dotfill &$20.8^{+1.0}_{-1.1}$&$38.3^{+4.1}_{-3.8}$&$121^{+13}_{-19}$\\

\smallskip\\\multicolumn{2}{l}{Wavelength Parameters:}&&TESS&\\
~~~~$u_{1}$\dotfill &\multicolumn{2}{l}{Linear limb-darkening coeff \dotfill }&$0.53^{+0.18}_{-0.23}$\\
~~~~$u_{2}$\dotfill &\multicolumn{2}{l}{Quadratic limb-darkening coeff \dotfill}&$-0.05^{+0.32}_{-0.23}$&\smallskip\\

&&14.98$\upmu$m&3.03$\upmu$m&4.51$\upmu$m\\
~~~~$u_{1}$\dotfill &Linear limb-darkening coeff \dotfill &$0.29^{+0.31}_{-0.21}$&$0.18^{+0.13}_{-0.12}$&$0.032^{+0.038}_{-0.023}$\\
~~~~$u_{2}$\dotfill &Quadratic limb-darkening coeff \dotfill &$0.29^{+0.31}_{-0.21}$&$0.25^{+0.18}_{-0.16}$&$0.043^{+0.053}_{-0.031}$\\
~~~~$A_T$\dotfill &Thermal emission from the planet (ppm)\dotfill &$173\pm17$&--&--\\
~~~~$\delta_{S}$\dotfill &Measured eclipse depth (ppm)\dotfill &$173\pm17$&--&--\\

\smallskip\\\multicolumn{2}{l}{Telescope Parameters:}&CARMENES&HARPS&HIRES\\
~~~~$\gamma_{\rm rel}$\dotfill &Relative RV Offset (m/s)\dotfill &$-1.84^{+0.86}_{-0.84}$&$-5.31\pm0.32$&$0.78^{+0.70}_{-0.71}$\\
~~~~$\sigma_J$\dotfill &RV Jitter (m/s)\dotfill &$1.83^{+1.2}_{-0.90}$&$1.86^{+0.29}_{-0.25}$&$3.02^{+0.71}_{-0.62}$\\
~~~~$\sigma_J^2$\dotfill &RV Jitter Variance \dotfill &$3.4^{+5.6}_{-2.5}$&$3.46^{+1.2}_{-0.87}$&$9.1^{+4.8}_{-3.4}$\smallskip\\

\smallskip\\\multicolumn{2}{l}{Telescope Parameters:}&PFSpost&PFSpre&UVES\\
~~~~$\gamma_{\rm rel}$\dotfill &Relative RV Offset (m/s)\dotfill &$-1.23^{+0.93}_{-0.90}$&$-2.1^{+1.2}_{-1.1}$&$1.13\pm0.67$\\
~~~~$\sigma_J$\dotfill &RV Jitter (m/s)\dotfill &$2.33^{+1.3}_{-0.89}$&$2.4^{+1.8}_{-1.2}$&$2.15^{+0.80}_{-0.77}$\\
~~~~$\sigma_J^2$\dotfill &RV Jitter Variance \dotfill &$5.4^{+7.7}_{-3.4}$&$5.6^{+12}_{-4.2}$&$4.6^{+4.1}_{-2.7}$\\

\end{longtable}
\tablefoot{
See Table~3 in \citet{Eastman2019} for a detailed description
of all parameters.
\tablefoottext{1}{This value ignores the systematic error and is
for reference only.}
\tablefoottext{2}{Time of conjunction is commonly reported as the
``transit time''.}
\tablefoottext{3}{\tjdtdb{} is the target's barycentric frame and
corrects for light-travel time.}
\tablefoottext{4}{Time of minimum projected separation is a more
correct ``transit time''.}
\tablefoottext{5}{Use this to model TTVs.}
\tablefoottext{6}{At the epoch that minimises the covariance between
$T_C$ and the period.}
\tablefoottext{7}{Use this to predict future transit times.}
\tablefoottext{8}{Assumes zero albedo and perfect heat redistribution.}
}
\clearpage

\section{Baseline-model fit parameters}
\label{app:fit_parameters}

\renewcommand{\arraystretch}{1.5}

\begin{table}[H]
  \caption{GJ~357\,b results for the 30\,min cutout: occultation depth, RMS, $\Delta$BIC, and fitted parameters. Values are medians with 16--84\% credible intervals.}
  \label{tab:gj357b_30min_all}
  \centering
  \setlength{\tabcolsep}{3pt}
  \resizebox{\textwidth}{!}{%
  \begin{tabular}{lccccccccc}
    \hline\hline
    Label
      & Occultation depth
      & RMS (ppm)
      & $\Delta$BIC
      & $a/R_\star$
      & $e$
      & $\omega$ (deg)
      & $i$ (deg)
      & $\Delta t_0$ (d)
      & $\Delta P$ (d) \\
    \hline

    Linear
      & $200.51^{+12.71}_{-12.67}$
      & 651
      & 0.00
      & $21.201^{+0.516}_{-0.503}$
      & $0.07880^{+0.02414}_{-0.02336}$
      & $82.081^{+2.025}_{-3.523}$
      & $89.399^{+0.208}_{-0.196}$
      & $-0.000009^{+0.000172}_{-0.000175}$
      & $0.00000034^{+0.00000613}_{-0.00000589}$ \\

    Linear + GP
      & $197.91^{+18.09}_{-19.22}$
      & 649
      & 29.20
      & $21.412^{+0.543}_{-0.525}$
      & $0.06916^{+0.02725}_{-0.02223}$
      & $81.180^{+2.775}_{-4.424}$
      & $89.342^{+0.223}_{-0.224}$
      & $-0.000000^{+0.000165}_{-0.000176}$
      & $0.00000080^{+0.00000611}_{-0.00000643}$ \\

    Lin + Exp
      & $203.24^{+12.92}_{-13.55}$
      & 652
      & 32.07
      & $21.212^{+0.543}_{-0.512}$
      & $0.07972^{+0.02509}_{-0.02218}$
      & $82.143^{+2.028}_{-3.149}$
      & $89.392^{+0.204}_{-0.189}$
      & $-0.000011^{+0.000179}_{-0.000179}$
      & $0.00000039^{+0.00000608}_{-0.00000631}$ \\

    Lin + Exp + GP
      & $203.49^{+22.41}_{-17.94}$
      & 649
      & 66.22
      & $21.421^{+0.624}_{-0.553}$
      & $0.07296^{+0.02820}_{-0.02511}$
      & $81.608^{+2.537}_{-5.691}$
      & $89.300^{+0.253}_{-0.253}$
      & $-0.000030^{+0.000183}_{-0.000154}$
      & $-0.00000077^{+0.00000613}_{-0.00000615}$ \\

    \hline
  \end{tabular}%
  }

  \tablefoot{
  $\Delta t_0$ and $\Delta P$ denote offsets from the adopted reference transit time and orbital period, respectively, and are given in days.
  }
\end{table}

\clearpage

\section{Corner Plot: Linear Fit with 30-Minute Cutout}
\label{app:corner_plot}

\begin{figure}[h]
    \centering
    \includegraphics[width=0.9\textwidth]{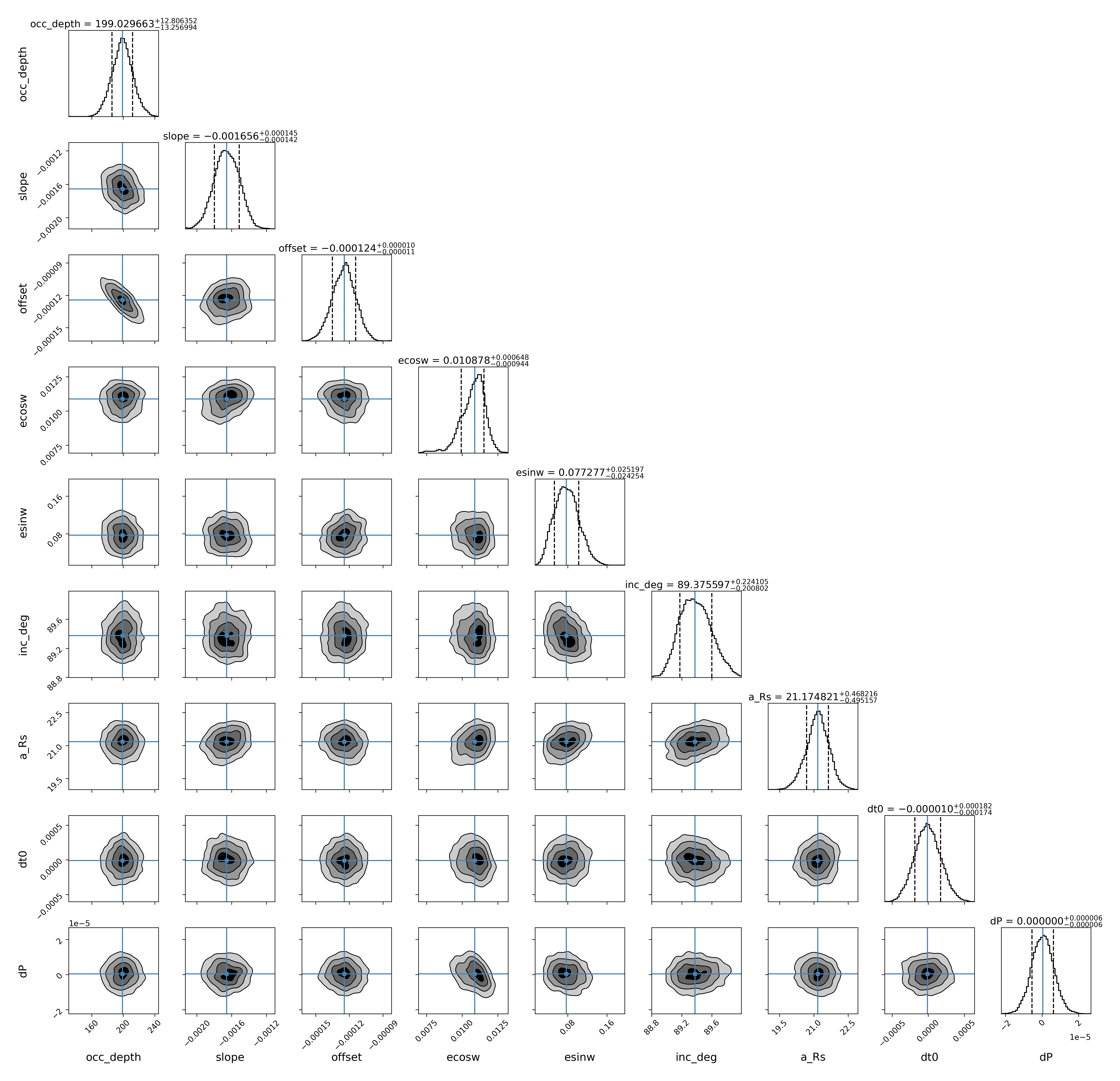}
    \caption{Corner plot showing posterior probability distributions from the MCMC fit to the GJ~357~b occultation light curve using a linear trend model with the first 30 minutes of data excluded. Diagonal panels show 1D marginalized posteriors with vertical dashed lines marking the 16th, 50th, and 84th percentiles. Off-diagonal panels show 2D joint posteriors with 68\% and 95\% confidence contours. Parameters include the occultation depth (ppm), linear trend slope and offset, orbital parameters ($e\cos\omega$, $e\sin\omega$, $i$, $a/R_\star$), and ephemeris corrections ($\Delta t_0$, $\Delta P$).}
    \label{fig:corner_fig}
\end{figure}

Figure~\ref{fig:corner_fig} presents the posterior probability distributions from our MCMC analysis of the GJ~357~b occultation light curve using a linear trend model with the first 30 minutes of data excluded. This cutout removes the initial thermal settling period of the detector, ensuring reliable photometric precision. The diagonal panels show the 1D marginalized posterior distributions for each fitted parameter, with vertical dashed lines indicating the 16th, 50th (median), and 84th percentiles. The off-diagonal panels display the 2D joint posterior distributions between parameter pairs, with contours enclosing 68\% and 95\% of the probability mass.

We fit for the occultation depth, linear trend parameters (slope and offset), orbital geometry ($e\cos\omega$, $e\sin\omega$, inclination, and scaled semi-major axis $a/R_\star$), and ephemeris corrections ($\Delta t_0$ and $\Delta P$). The tight, Gaussian-like posteriors for most parameters indicate well-constrained solutions with minimal parameter degeneracies. Notable correlations appear between the occultation depth and the linear slope, as well as between $e\cos\omega$ and $e\sin\omega$, reflecting the geometric coupling between eccentricity and argument of periastron. The timing offsets $\Delta t_0$ and $\Delta P$ are consistent with zero, validating the accuracy of the ExoFASTv2 ephemeris used as our prior.

\clearpage

\section{Detailed Model Families}
\label{app:model_families}

\begin{figure}[h]
    \centering
    \includegraphics[width=0.9\textwidth]{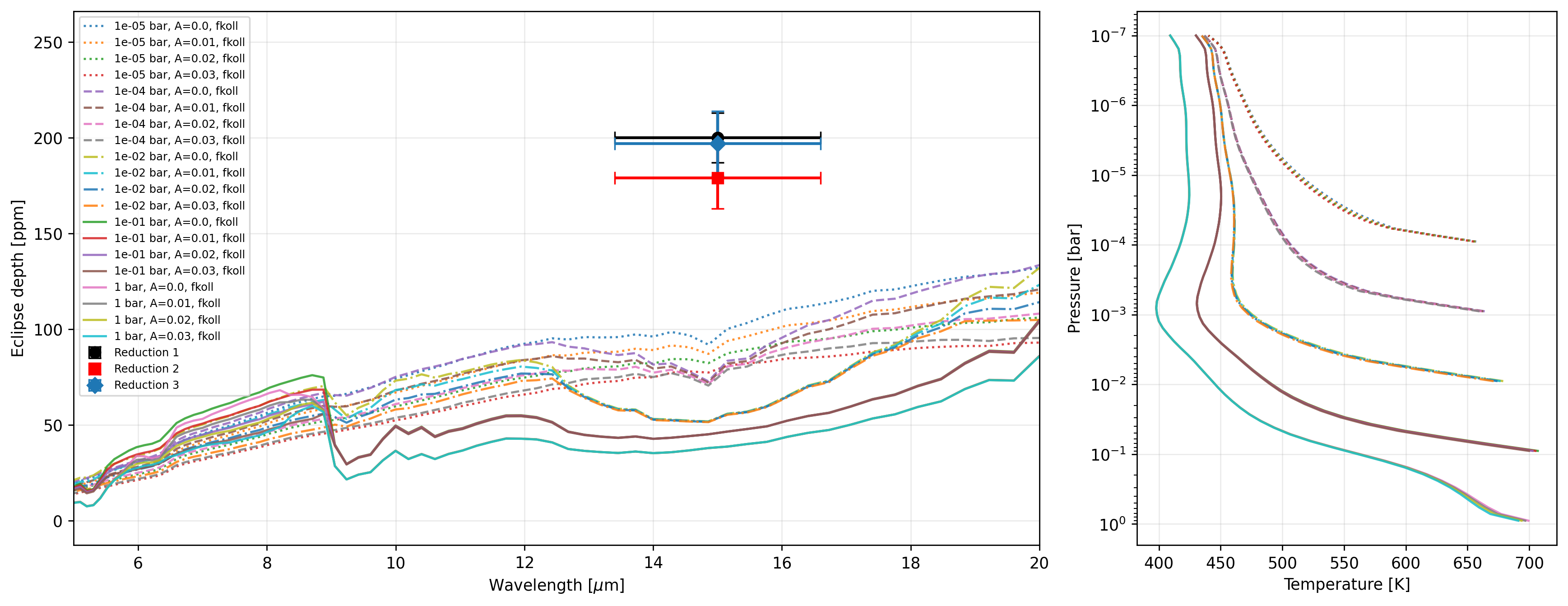}
    \caption{Pure CO$_2$ atmosphere models.}
    \label{fig:family_co2}
\end{figure}

\begin{figure}[h]
    \centering
    \includegraphics[width=0.9\textwidth]{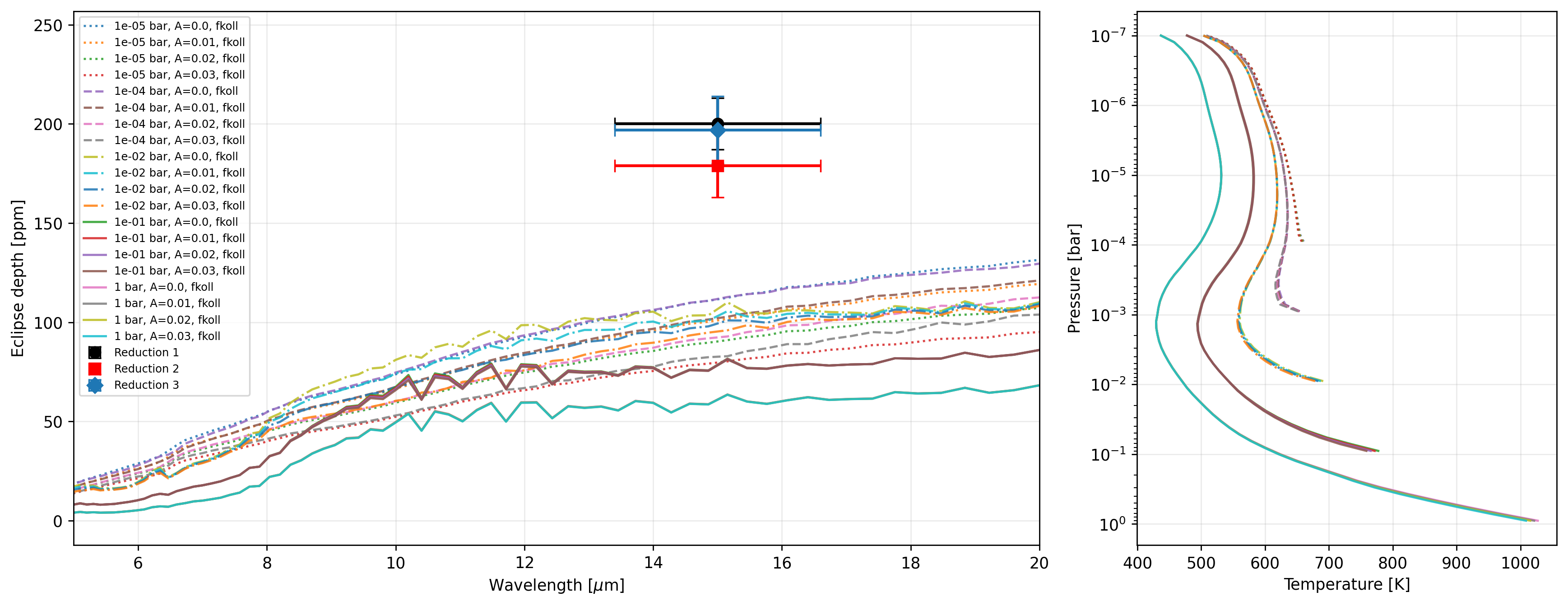}
    \caption{Pure H$_2$O atmosphere models.}
    \label{fig:family_h2o}
\end{figure}

\begin{figure}[h]
    \centering
    \includegraphics[width=0.9\textwidth]{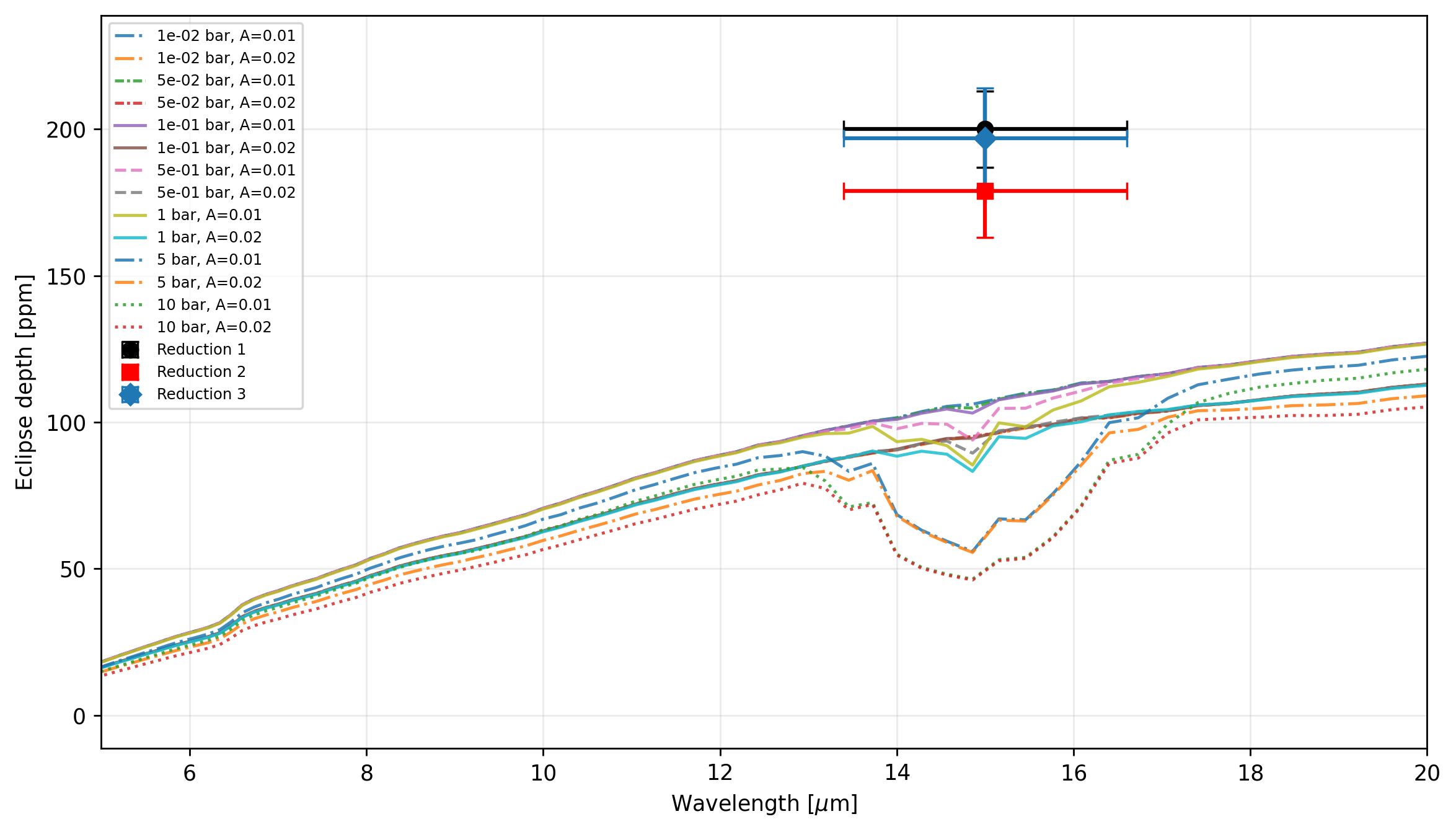}
    \caption{N$_2$ atmosphere models containing 1~ppm CO$_2$.}
    \label{fig:family_n2_co2_1ppm}
\end{figure}

\begin{figure}[h]
    \centering
    \includegraphics[width=0.9\textwidth]{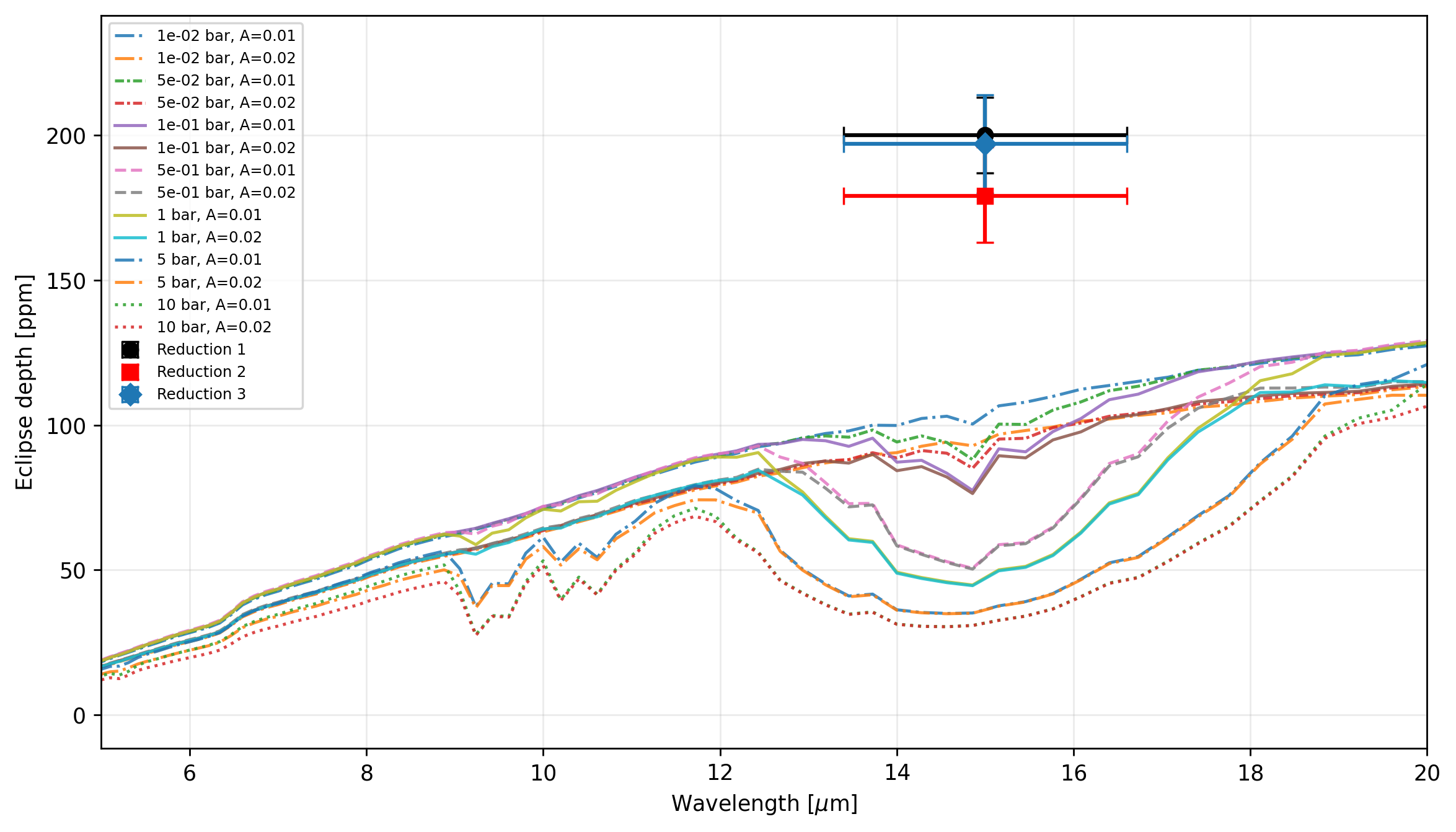}
    \caption{N$_2$ atmosphere models containing 100~ppm CO$_2$.}
    \label{fig:family_n2_co2_100ppm}
\end{figure}

\begin{figure}[h]
    \centering
    \includegraphics[width=0.9\textwidth]{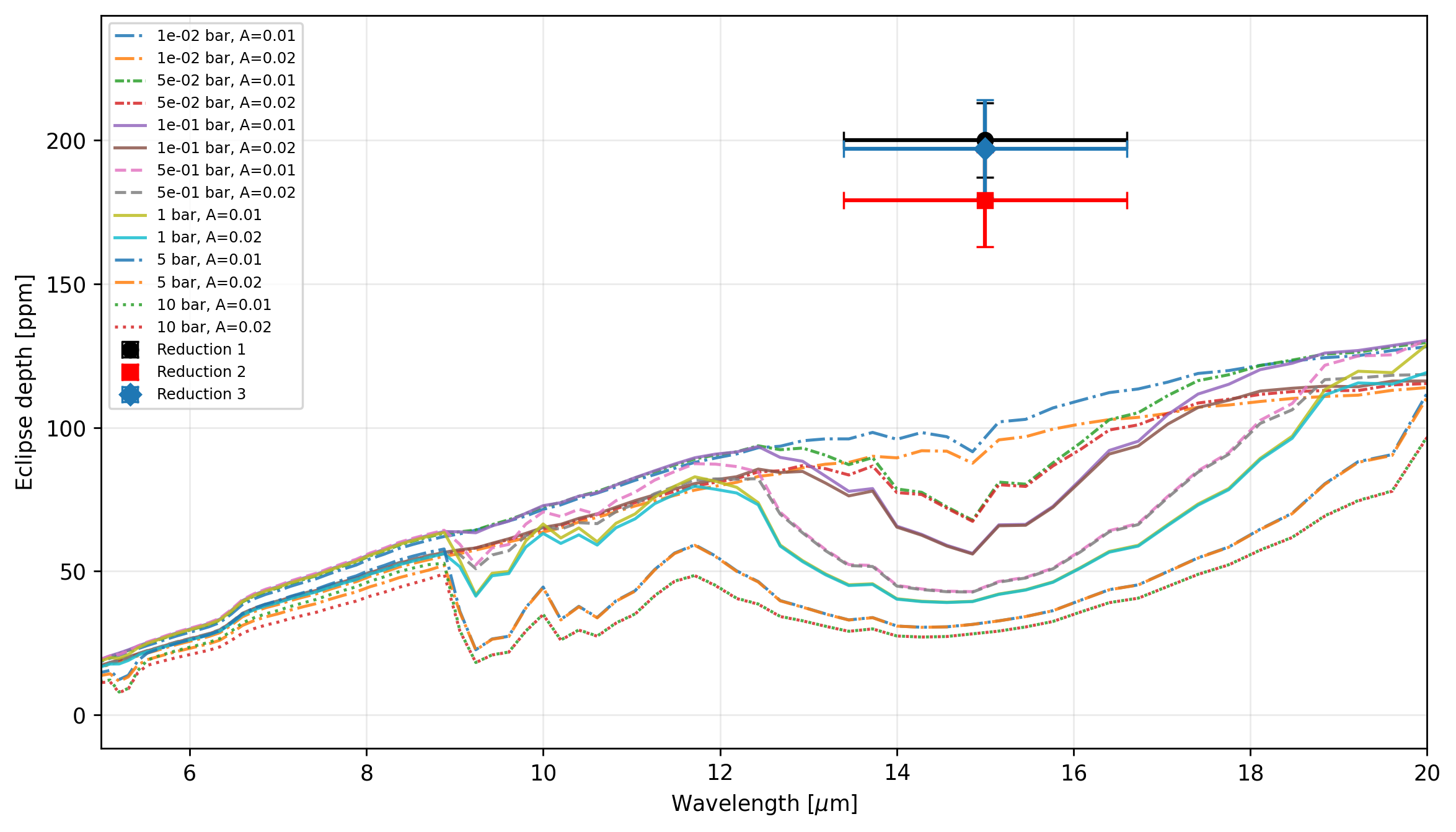}
    \caption{N$_2$ atmosphere models containing 1000~ppm CO$_2$.}
    \label{fig:family_n2_co2_1000ppm}
\end{figure}

\begin{figure}[h]
    \centering
    \includegraphics[width=0.9\textwidth]{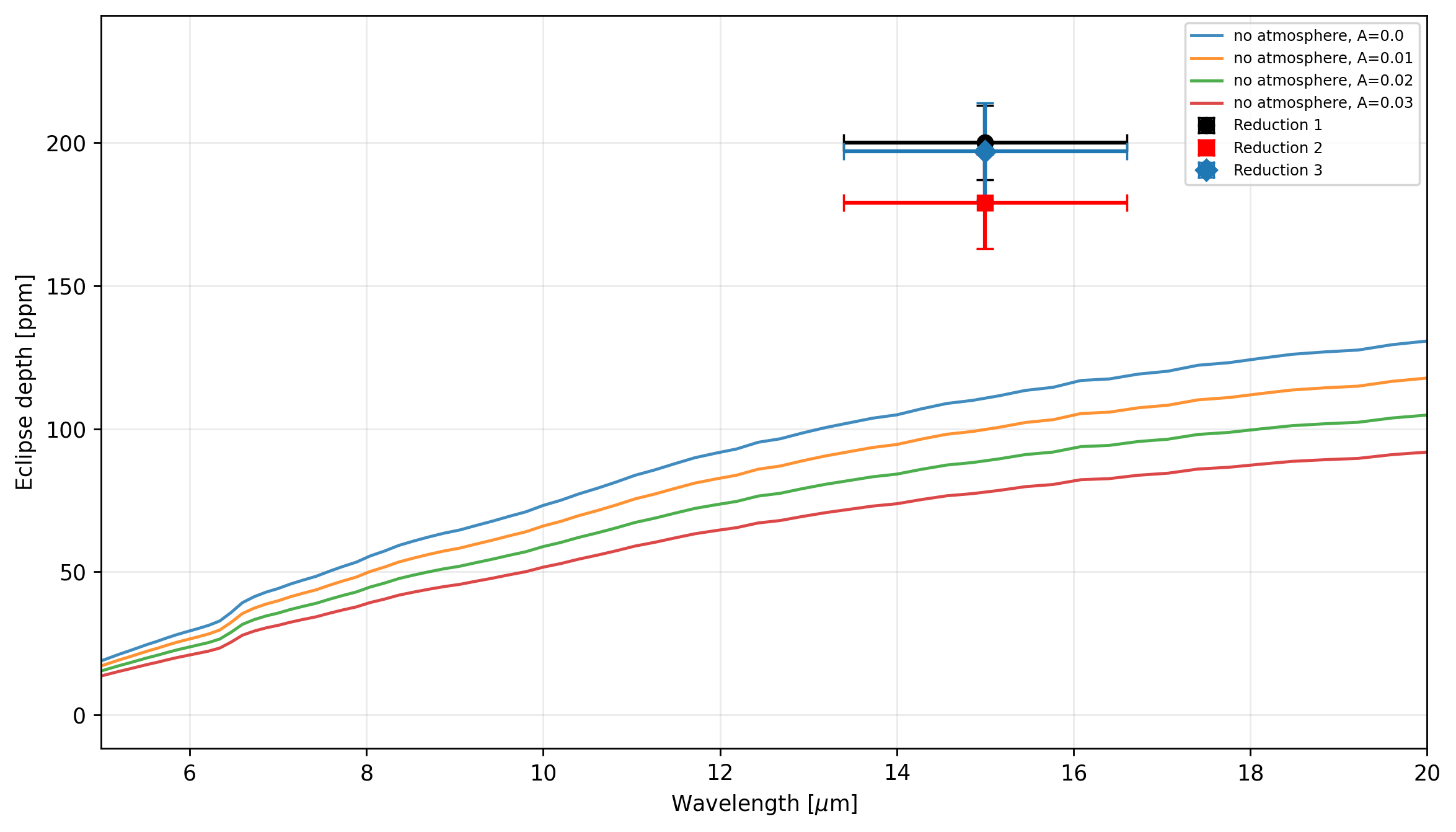}
    \caption{Zero-albedo no-atmosphere model shown for comparison.}
    \label{fig:family_noatmos}
\end{figure}

Figures~\ref{fig:family_co2}--\ref{fig:family_noatmos} show detailed comparisons for the atmospheric model families considered in this work. These include pure CO$_2$, pure H$_2$O, N$_2$ atmospheres containing 1, 100, and 1000~ppm CO$_2$, and the no-atmosphere model. Line styles encode surface pressure as described in the main text. Here, $A$ denotes the Bond albedo.

\clearpage
\section{SPHINX stellar models varying C/O ratio}\label{app:sphinx}

As part of our exploration in comparing flux-calibrated measured data to stellar spectral models, we present a series of SPHINX stellar models \citep{Iyer2023,Iyer2026} with C/O ratios ranging from 0.3 to 0.9. We compare the models to flux calibrated measurements of the host star from NIRISS/SOSS \citep[GO 1201, PI Lafreniere;][]{Taylor_2025}, NIRSpec G395H \citep[GO 2512, PI Batalha;][]{AdamsRedai2025}, and MIRI/F1500W (GO 3730, PI Diamond-Lowe; this work) observations. We adopt the SPHINX model with C/O=0.3 as our nominal stellar model. 

\begin{figure}[h]
    \centering
    \includegraphics[width=\textwidth]{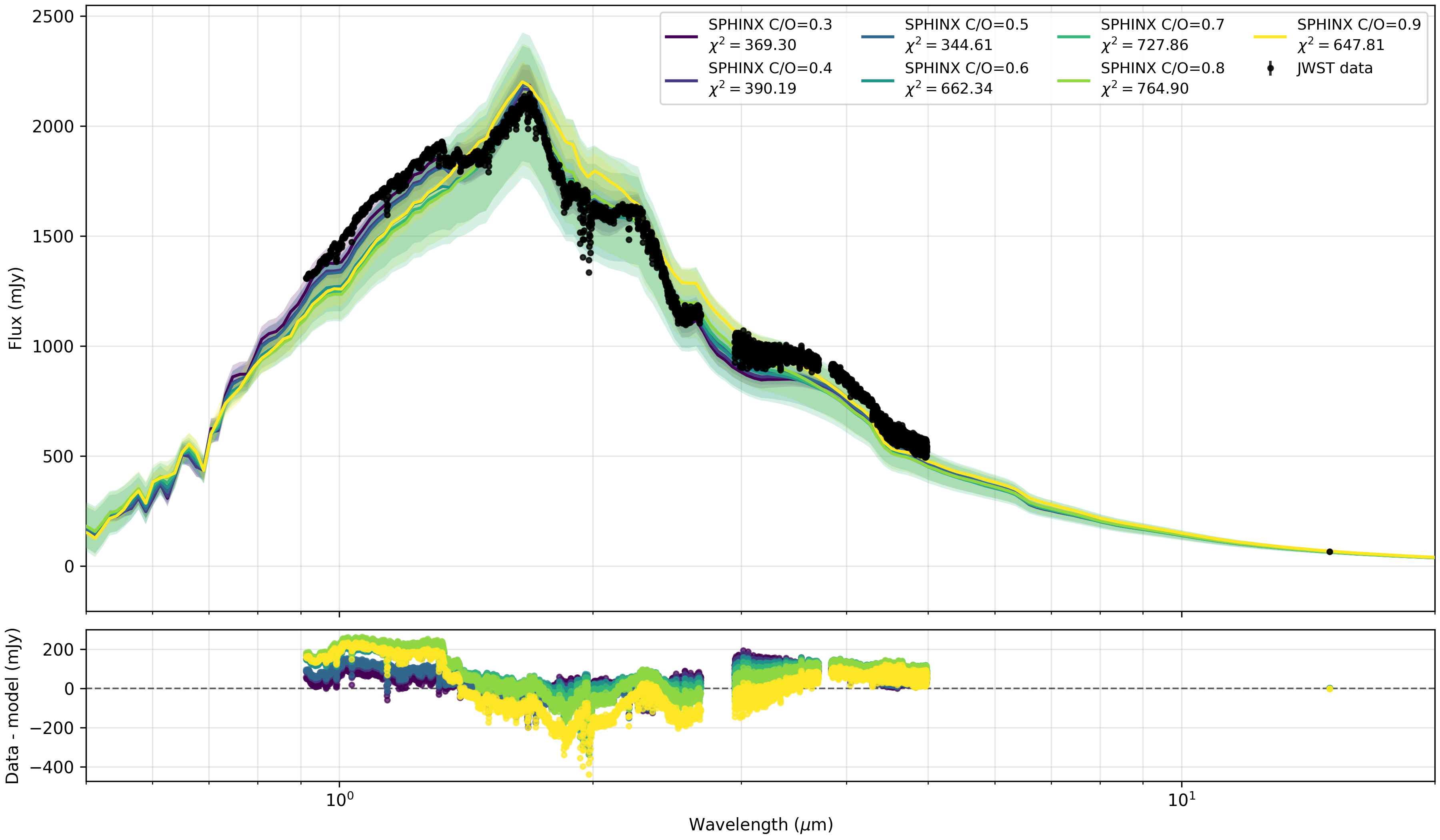}
    \caption{SPHINX models with C/O ranging from 0.3 to 0.9 for the derived GJ 357 stellar parameters. The $\chi^2$ value reflects the comparison between the data and the models; it is dominated by the NIRISS SOSS data as there are more data points. We adopt the SPHINX model with C/O=0.3 as our nominal stellar model. }
    \label{fig:sphinx_co_grid}
\end{figure}

\end{document}